\documentclass[letterpaper, 10 pt, conference]{ieeeconf}

\IEEEoverridecommandlockouts 

\usepackage{graphics} 
\usepackage{epsfig} 
\usepackage{amsmath} 
\usepackage{amssymb} 
\usepackage{amsfonts}
\usepackage{mathtools}
\usepackage{caption}
\usepackage{algorithm}
\usepackage{array}
\usepackage{textcomp}
\usepackage{stfloats}
\usepackage{url}
\usepackage{verbatim}
\usepackage{graphicx}
\usepackage{cite}
\usepackage[inkscapeformat=png]{svg}
\usepackage{subcaption}
\usepackage[font={scriptsize}]{caption}
\usepackage[section]{placeins}
\usepackage{float}
\usepackage{booktabs}
\usepackage{makecell}
\usepackage{blindtext}
\usepackage{comment}
\usepackage{algpseudocode}
\usepackage{tablefootnote}
\usepackage{footnotehyper}
\usepackage[colorlinks,linkcolor=blue]{hyperref}
\usepackage{xcolor}

\newtheorem{assumption}{Assumption}

\graphicspath{{./figures/}}
\usepackage{balance}
\usepackage{multirow}
\usepackage{color,soul}
\usepackage{framed}
\usepackage{wrapfig}
\usepackage{import}
\usepackage{arydshln}
\usepackage{multicol}
\usepackage{hyperref}

\hypersetup{
 colorlinks=true,
 linkcolor=black,
 citecolor=black,
 filecolor=magenta, 
 urlcolor=blue,
 }

\newcommand{\reals}{\mathbb{R}}

\definecolor{RevBlue}{HTML}{0072B2}
\definecolor{RevVermilion}{HTML}{D55E00}
\definecolor{RevGreen}{HTML}{009E73}
\definecolor{RevOrange}{HTML}{E69F00}
\definecolor{RevPurple}{HTML}{CC79A7}

\newcommand{\URL}[1]{{\color{olive}\text{Supplemental Video:} #1}}

\DeclarePairedDelimiter\norm{\lVert}{\rVert}%

\title{\Large
Safe On-Orbit Dislodging of Deployable Structures via Robust Adaptive MPC 

}
\author{Longsen Gao$^{1,*}$, Claus Danielson, Andrew Kwas, and Rafael Fierro
\thanks{Longsen Gao and Rafael Fierro are with the Department of Electrical and Computer Engineering, The University of New Mexico, Albuquerque, NM 87131, USA. {\tt\small \{lgao1, rfierro\}@unm.edu}}
\thanks{Claus Danielson is with the Department of Mechanical Engineering, The University of New Mexico, Albuquerque, NM 87131, USA. {\tt\small \{cdanielson\}@unm.edu}}
\thanks{Andrew Kwas is with the Department of Electrical and Computer Engineering, The University of New Mexico and with Northrop Grumman Corporation, USA.{\tt\small \{andrew.kwas\}@ngc.com}}
}
\begin{document}

\maketitle

\begin{abstract}
This paper proposes a novel robust adaptive model predictive controller for on-orbit dislodging. 
We study orbit dislodging where a servicing spacecraft uses a robotic arm to free a jammed and unactuated solar panel mounted on a hybrid hinge that acts as a time-varying client on a space station.
Our method couples online set-membership identification with a robust adaptive MPC to enforce safety under bounded disturbances. The controller explicitly balances exploration to excite the system and shrink uncertainty and exploitation to improve control performance through a dual-mode cost.
The feasibility of the developed robust adaptive MPC method is also examined through dislodging simulations and hardware experiments in freefall and terrestrial laboratory environments, respectively. In addition, the advantages of our method are shown through comparison experiments with several state-of-the-art control schemes for both accuracy of parameter estimation and control performance. 
\end{abstract}

\section*{NOMENCLATURE}
\begin{description}
    \item[$M$] Inertia matrix.
    \item[$C$] Coriolis and centripetal torque term.
    \item[$\mathcal{K}$] Stiffness parameter.
    \item[$\mathcal{D}$] Grasp factor related to contact geometry.
    \item[$A$] State matrix of the linear parametric model.
    \item[$A_i$] Affine decomposition matrix of $A(\rho_k)$.
    \item[$B$] Input matrix of the linear parametric model.
    \item[$B_i$] Affine decomposition matrix of $B(\rho_k)$.
    \item[$\mathcal P$] Admissible parameter polytope.
    \item[$\mathbb{W}$] Bounded disturbance set.
    \item[$H_\rho$] Polytope matrices defining $\mathcal P$.
    \item[$H_w$] Polytope description of $\mathbb{W}$.
    \item[$\mathcal{U}$] Input constraint set.
    \item[$\mathcal{X}$] State constraint set.
    \item[$\mathbb{Z}$] Combined state–input constraint set.
    \item[$\tau_f$] Coulomb friction torque.
    \item[$\mathbf{K}$] Stabilizing feedback gain used in the tube policy.
    \item[$\mathbf{A}_\text{cl}$] Closed-loop map operator.
    \item[$\Gamma_{j,\varsigma|k}$] Nonnegative multipliers certifying tube constraints.
    \item[$\mathbb{X}_{T}$] Terminal set with robust positive invariance in tube variables.
    \item[{\small $Q{,}R{,}\Upsilon$}] Weighting matrix in the cost.
    \item[$\varOmega$] Collection of decision variables.
    \item[$\lambda$] Scaling factor used with $\mathbb{X}_{T}$ in the terminal condition.
    \item[$J_N$] Finite-horizon objective.
    \item[$\mu$] Friction coefficient.
    \item[$r_e$] Effective rotor radius.
    \item[$\epsilon$] Small positive smoothing constant.
    \item[$\tau_{m,k}$] External torque disturbance applied by the manipulator.
    \item[$\delta_\tau$] Bound on the external torque disturbance.
    \item[$\varsigma$] Step index along the prediction horizon.
    \item[$\alpha_k$] Angle between gripper direction and rod axis at time $k$.
    \item[$d_g$] Dislodging contact position measured from the hinge axis.
    \item[$r_\alpha$] Wrench direction vector used in the torque transfer term.
    \item[$\boldsymbol{\rho}$] Parameter vector.
    \item[$\zeta_k$] Viscous damping parameter.
    \item[$\phi_\rho$] Bound on the rate of change of $\boldsymbol{\rho}_k$.
    \item[$v_{\varsigma|k}$] Auxiliary input in the tube policy.
    \item[$c_{\varsigma|k}$] Auxiliary input used in the convex reformulation.
    \item[$\ell$] Stage and terminal cost function.
\end{description}

\section{Introduction}\label{sec:intro}
On-orbit failures have frequently occurred ever since the first artificial satellite was launched into space~\cite{emme1966part,tafazoli2009study,flores2014review,luu2021review} due to multiple factors including orbital debris~\cite{putzar2008vulnerability}, thermal cycling and temperature extremes~\cite{putz2020mechanical}, radiation-induced degradation and surface charging~\cite{buitrago2024statistical}, vacuum-induced cold welding and stiction~\cite{merstallinger2021cold}, and electromagnetic interference~\cite{de2022simulation}. These failures have caused countless economic and scientific losses, which have become an enormous challenge in the past decades. On-orbit servicing leveraging autonomous robotics systems is an increasingly popular research topic for rescuing on-orbit failure~\cite{davis2019orbit,oestreich2021orbit,virgili2019convex,parikh2025safe}. 
Additionally, servicing missions confront two fundamental challenges. Model uncertainty from partially known and time-varying client parameters and failure modes degrades prediction and planning, while operational limits including limited sensing~\cite{boning2006identification}, tight actuation bounds~\cite{ma2016adaptive}, and strict real-time requirements shrink the feasible set of safe control actions~\cite{beaumet2011feasibility}.

Dislodging jammed components, where off-nominal friction temporarily raises the torque demand above the joint’s nominal requirement, remains a common on-orbit maintenance task~\cite{laing2009space,rivera2021study,jammedSolar}. It involves safely manipulating objects that may be stuck or restrained due to various factors, such as microgravity~\cite{higuchi2002unexpected}, stiction~\cite{rivera2021study}, or mechanical failure~\cite{tafazoli2009study}. Dislodging a jammed component often requires applying significant, sudden force, which could potentially damage a delicate component. Few studies have considered using on-orbit robotics systems to address the problem. Previous work in~\cite{gao2023autonomous} presented an adaptive control algorithm using a multi-robot system in a dislodging task for a solar panel. However, the algorithm did not consider safety constraints, which can lead to damage of both the servicer and client spacecraft. 
The risk of damage is exacerbated during the transient period when the adaptive controller is learning the system parameters. Although the adaptive controller is guaranteed to eventually learn the correct parameter, during the learning process, the adaptive controller can exert excessive force to dislodge the jammed component, potentially causing damage. 

Online parameter estimation can be leveraged to address the challenge of uncertainty during servicing missions to guarantee safety. Adaptive control for updating controller parameters using real-time measurement data has been extensively studied and gained popularity in recent decades~\cite{xu1993adaptive,ulrich2010modified,yu2019robust}. 
Traditional Model Reference Adaptive Control (MRAC)~\cite{nguyen2018model} updates controller parameters to track a reference model but is sensitive to unmodeled dynamics and does not natively enforce state or input constraints
, which can be unreliable in practical implementation due to unmodeled dynamics. Failure to enforce constraints on both the manipulator and manipulated object may lead to component damage during the servicing~\cite{enomoto2023delicate}, which can cause significant costs and lost time due to the need to replace aerospace components via relaunch.

Model predictive control (MPC) is widely adopted in terrestrial industries for explicit handling of constraints and for stability assurance, while within spacecraft guidance and control it is an limited but worthy of development and expanding research area with selective applications based on its strong safety and reliability. In~\cite{lorenzen2017adaptive}, an adaptive MPC algorithm for linear systems was introduced using a recursive approach, and its extension in~\cite{kohler2021robust} employed a tube-based framework to handle parametric uncertainty and additive disturbances in nonlinear systems. However, these methods rely on passive identification, preventing full exploitation of concurrent adaptation and control. A learning-based MPC in~\cite{aboudonia2024adaptive} used set-membership identification composed of two online phases: an adaptation phase with rigid tube-based robust MPC and a learning phase for uncertainty estimation, but only considers constant parameters, making it unsuitable for time-varying applications such as those found in aerospace. To address these limitations, this paper considers an adaptive controller that integrates set-membership identification to iteratively refine parameter bounds and employs robust MPC to ensure constraint enforcement and stability under worst-case conditions, ultimately improving performance as uncertainty estimates become less conservative.

In this paper, we propose a novel robust adaptive MPC with set-membership parameter estimation for the problem of safe dislodging. We demonstrate its implementation on the servicer during a dislodging task based on the client. The main contributions are summarized:
\begin{itemize}

    \item Our robust adaptive MPC method provides robust guarantees on constraint enforcement despite the parametric uncertainty of the client and unpredictable failure modes.
    \item We derive a state-space representation of the hinge model and implement the proposed MPC for the dislodging task. Comparative studies against two baseline methods evaluate feasibility and performance.
        
    \item We develop a novel cost function that incorporates the time-varying parameter set with control input and state to improve the performance of parameter estimation during the control process. 
\end{itemize}


The robust adaptive MPC in this work leverages dual-mode control~\cite{wesselowski2003dual} by incorporating time-varying parameter estimates into the MPC cost, balancing exploration which increases persistency of excitation to reduce uncertainty and exploitation which enhances control performance. Comparisons with a PID controller, adaptive control~\cite{gao2023autonomous}, and a state-of-the-art AMPC~\cite{aboudonia2024adaptive} demonstrate the superior performance and parameter estimation for time-varying systems. Moreover, the dual-state tube ensures safe parameter estimation within the initial set, remaining robust against future uncertainties.
\begin{figure}[!t]
\centering
\captionsetup{font=footnotesize}
\includegraphics[width=0.47\textwidth]{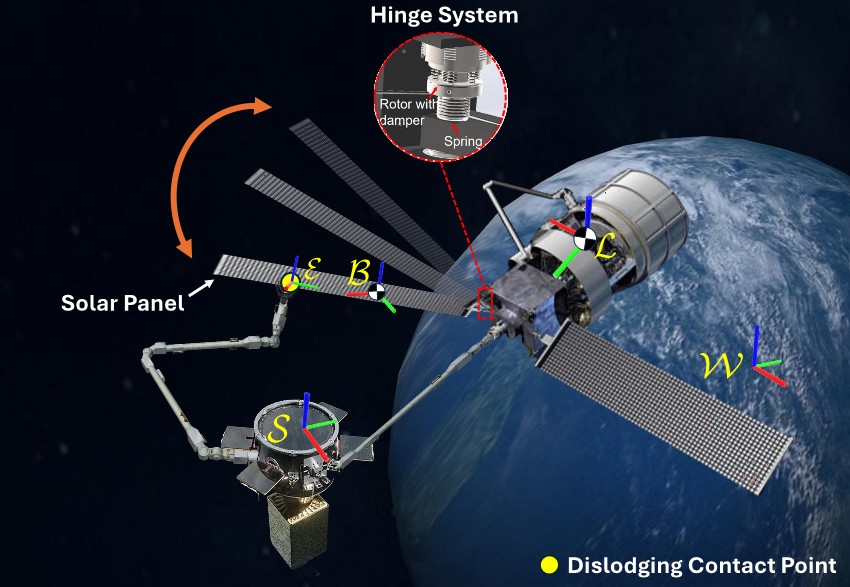}
\caption{A \textit{servicer} docked with a space station dislodges an unactuated solar panel with hinge via its robot arm in space. \URL{\href{http://tiny.cc/tcst2025unm}{http://tiny.cc/tcst2025unm}} }
\label{fig::diagram_detail}
\end{figure}
\subsubsection*{Notation}
The sets of real numbers and positive real numbers are denoted by $ \mathbb{R} $ and $ \mathbb{R}_+ $, respectively. The sequence of integers from $ n_1 $ to $ n_2 $ is represented by $\{n_1,n_2\}_\mathbb{N}=\{n \in \mathbb{N} \mid n_1 \leq n \leq n_2\}$. For a vector $\mathbf{p}$, $\mathbf{p}{[i]}$ refers to its $i\text{-th}$ element. The $i\text{-th}$ row of a matrix $\mathbf{A}$ is denoted by $ [\mathbf{A}]_{i} $. The Minkowski sum of two sets $\mathcal{A}$ and $\mathcal{B}$ is denoted by $ \mathcal{A} \oplus \mathcal{B} $, and the pontryagin set difference of two sets can be denoted as $\mathcal{A} \ominus \mathcal{B}$. $\boldsymbol{\gamma} \in \reals^N$ denotes a column vector with $N$ rows that each row contains a scalar value $\gamma$. The convex hull of the elements of a set S is represented by $\textbf{co}$\{S\}. We use $\mathbf{u}_k$ for the (real, measured) vector at time $k$ and $\mathbf{x}_{\varsigma|k}$ for the vector predicted $\varsigma$ steps ahead at time $k$. The estimated term is denoted with a ($ \hat{~} $) on the top and the upper bound of a variable using ($\  \bar{} \ $) on the top. 

\section{On-Orbit Dislodging Problem}

In this section, we define the dislodging problem for an unactuated jammed solar panel, preventing it from unfolding through its passive actuation system. Considering both unknown parameters and control input from the unactuated jammed solar panel and manipulator, respectively, with their constraints during dislodging, the objective is to keep safety not only for the control action but also for parameter estimation.

\subsection{Client and Servicer Agents}

Let's define the servicer and client agents and their interaction in the dislodging problem. 

\textbf{Client} The client is a free-flying resident space object (RSO) with a jamming solar panel. We assume the actuation system of the solar panel on its hinge is jamming and cannot unfold the solar panel to the desired position. The hinge is stronger than the truss on the solar panel. The solar panel is also jammed in a random position. 

\textbf{Servicer}
The servicer is a multi-functional spacecraft that can perform highly complicated repairing and servicing tasks. We assume that the servicer includes $1$ robotic arm with a gripper that can hold the solar panel to dislodge it into a specific position to unfold it.

We assume the frame $\mathcal{L}$ is a a standard Local-Vertical Local-Horizontal(LVLH) frame placed on the center of mass(CoM) of the space station of the \textit{Client}. $\mathcal{W}$ denotes the earth-fixed frame, which should be located at the center of the Earth, as shown in Fig.~\ref{fig::diagram_detail}. Considering the client is a free-flying aerospace system that can adjust it motion to decouple the dynamics from the solar panel during dislodging automatically by its inner stabilized system. This assumption holds during the whole process of dislodging. 

\begin{figure*}[!t]
\centering
\captionsetup{font=footnotesize}
\includegraphics[width=0.85\textwidth]{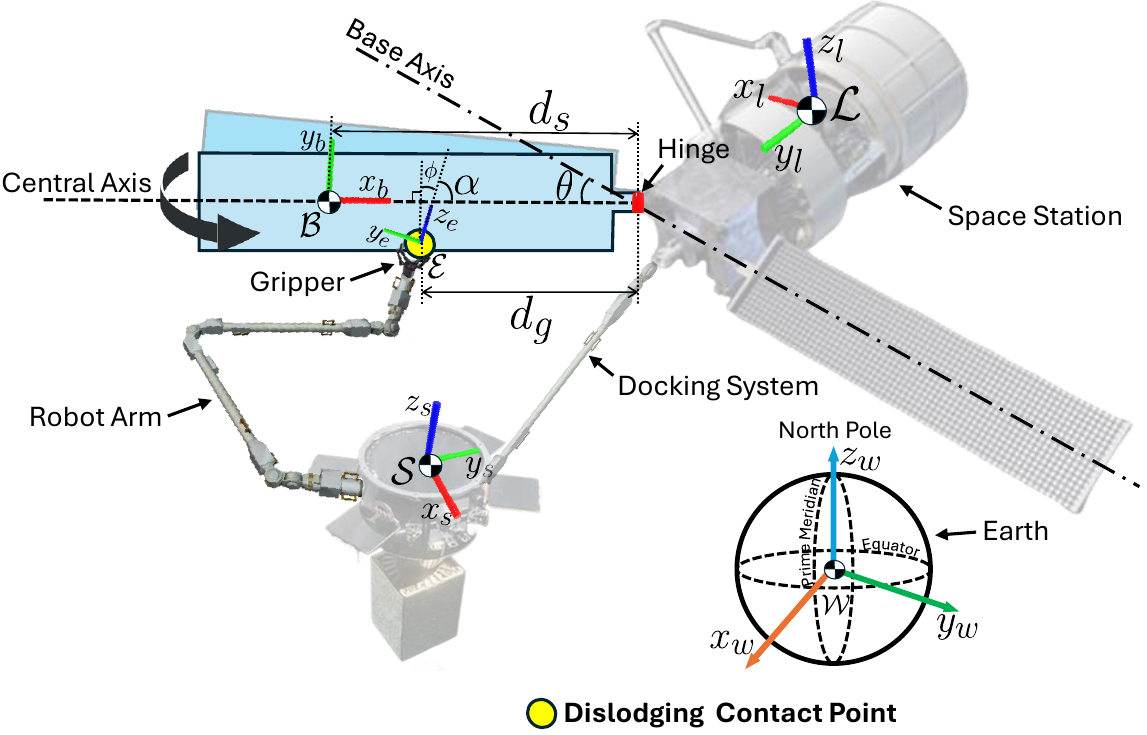}
\caption{Diagram of the dislodging process that one servicer docks with a space station using its docking system and then dislodges a jamming solar panel that is unactuated via its manipulator in space.}
\label{fig::diagram_fig1}
\end{figure*}

Furthermore, we consider the client as a time-varying system in which the external environment may affect its stiffness properties, e.g., temperature, macro-gravity,  vacuum conditions, radiation, etc.,~\cite{tribble2020space}. Also, we assume the dislodging contact location and orientation are ambiguous; the robot arm may not always be perpendicular to the rod, which means the $\alpha$ is not always $90^\circ$ and should also be time-varying as shown in Fig.~\ref{fig::diagram_fig1}. Note that the docking system is a rigid connector, as shown in Fig.~\ref{fig::diagram_fig1}, which can ensure the \textit{Servicer} keeps relative static with the space station during dislogding mission, and there is no relative decoupling dynamics needed. Hence, estimating unknown time-varying parameters for both the jammed component and grasping information during the dislodging under constraints becomes the biggest challenge in this task. We will pose this problem by dislodging the hinge from its initial position $\theta_0$ to a desired position $\theta^*$ with the calculated force applied by the servicer. The CoM is located at the position with distance from the pivot axis of the revolute joint as $d_s$. The dislodging contact position relative to the revolute joint is $d_g$. The angle between the z-axis of the end-effector (EE) of the robot arm on \textit{Servicer} and the central axis of the rod denotes $\alpha$, and the angle between the z-axis of EE and its perpendicualr axis relative to the central axis of the solar panel denotes $\phi$ in which $\phi = 90^\circ - \alpha$ as shown in Fig.~\ref{fig::diagram_fig1}.

\subsection{Dislodging Dynamics}

In this section, we model the dynamics of the client which is unactuated and to be dislodged by the servicer. The functioning-well solar panel is modeled as an unactuated hybrid hinge~\cite{fang2012shape} system whose dynamics can be modeled by the Euler-Lagrange equation in joint space as
\begin{equation} \label{eq::Dynamics_euler}
 \begin{aligned}
    \tau_h= M {{\ddot \theta}_k}
    + \Lambda \dot{\theta}_k
 + \mathcal{K}_k \theta_k + \tau_f(\dot \theta_k) + \tau_{m_k},
 \end{aligned} 
\end{equation}
where $\tau_h = \mathbf{u}_k^\top \cdot \mathbf{r}_\alpha \times d_g \in \reals_{+}$ denotes the torque applied on the hinge on the \textit{client} by the \textit{servicer} during dislodging, in which $\mathbf{r}_\alpha = \left[0, -\cos \alpha, \sin \alpha, 0, 0, 0 \right]^\top \in \reals^6$ denotes the rotation vector for the EE of the robot arm on servicer relative to the perpendicular to the client; $ \mathbf{u}_k {=} \begin{bmatrix}
 \mathbf{f}_k,\boldsymbol{\tau}_k
\end{bmatrix}^\top {\in} \  \reals^6$ in which $\mathbf{f}_k {=} \left[{f_x}_k, {f_y}_k, {f_z}_k \right]^\top {\in} \reals^3$ and $\boldsymbol{\tau}_k {=} \left[{\tau_x}_k, {\tau_y}_k, {\tau_z}_k \right]^\top {\in} \ \reals^3$ denote the force and torque applied by the EE of the robot arm on the \textit{servicer} to the dislodging contact point, respectively. $\theta_k$ denotes the angle of the hinge; $M \in \reals_+$ denotes the mass inertia of the hinge. $\Lambda = C_k + \mathcal{\zeta}_k $ in which $C \in \reals_+$ is the Coriolis and centripetal torque of the solar panel with friction torque limiter\protect\footnotemark[1]
\footnotetext[1]{\href{https://www.machiii.com/resources/how-our-products-work/how-it-works-mechanical-friction-torque-limiter/}{Structure of friction torque limiter}, by MACH III CLUTCH.}
and $ \mathcal{\zeta}_k  \in \reals_+$ is the viscous friction coefficient of the hybrid hinge.. The spring-loaded hinge has stiffness $\mathcal{K}_k$ and viscous friction coefficient $\zeta_k$. The term $$\tau_f(\dot \theta) = -\mu \cdot r_e \cdot \tfrac{\dot \theta}{\left| \dot \theta \right| + \epsilon}$$ is the Coulomb friction of the hybrid hinge based on the dataset\protect\footnotemark[2] we collected with material-dependent friction coefficient $\mu$; $r_e$ denote the effective radius of the rotor and $\epsilon \rightarrow 0 \in \reals_+$ to avoid the denominator to be zero. $\tau_{m_k} \in \{\tau_m \in \mathbb{R}: |\tau_m| \leq \delta_\tau \}$ denotes the external disturbance by the manipulator which is bounded through a constant value $\delta_\tau \in \reals_+$ that affects the torque of the hinge. Note that this hybrid hinge is a torsional design, where a coaxial helical spring produces a restoring torque modeled by a linear stiffness law.
\footnotetext[2]{Supported material: {\href{https://huggingface.co/datasets/gaolongsen/RobustAdaptiveMPC_TCST}{Hinge Dataset}.}}

\subsection{Parametric Uncertainty}

The dynamics~\eqref{eq::Dynamics_euler} have parametric uncertainty. Specifically, the parameter vector $\boldsymbol{\rho}_k  = [\mathcal{K}_k, \zeta_k, \mathcal{D}_k]^\top \in \reals^{3}$ is an unknown and time-varying vector where $\mathcal{D}_k = d_g \cdot \sin(\alpha_k)$ denotes the relative grasping positions of the gripper of the robot arm on the servicer; $\boldsymbol{\rho}^*_k$ is the true value of $\boldsymbol{\rho}_k$. 
We assume that at each time $k$ the parameter $\boldsymbol{\rho}_k$ belong to the known bounded polytope
\begin{equation}\label{eq:ParameterBounds}
    \mathcal{P} := \{\boldsymbol{\rho}_k \in \mathbb{R}^3 | \mathbf{H}_{\rho} \boldsymbol{\rho}_k \le \boldsymbol{\delta}_{\rho} \},
\end{equation}
where $ \boldsymbol{\rho}_k^* \in \mathcal{P}$ and $\mathbf{H}_{\rho} \in \mathbb{R}^{n_p \times 3}$ in which $n_p$ denotes the number of the hyperplanes used to define the known bounded polytope $\mathcal{P}$. 
Furthermore, we assume the change in parameters is bounded. Specifically, we assume there exists a value $\varphi_\rho \in \reals_+$ that satisfies $\boldsymbol{\rho}_k^* \in \mathcal{P}$ for $k \in \mathbb{N}$ that
\begin{equation}
\label{eq:parameter-velocity}
    \left\|\boldsymbol{\rho}_{k+1}^*-\boldsymbol{\rho}_k^*\right\| \leq \varphi_\rho.
\end{equation}

We thus clarify that the system state $\boldsymbol{\theta}_k$ describes only the hinge angle, while the off-pointing geometry of the \textit{Servicer}'s EE $\alpha_k$ is treated as an unknown, time-varying parameter handled within the estimation vector $\boldsymbol{\rho}_k$.


\subsection{Safe Dislodging Constraints}

To prevent damaging both the manipulator on the servicer and the jammed component during dislodging, we first need to enforce the input constraints
\begin{subequations}
\label{eq:constraint}
\begin{align}
\label{eq:input_constraint}
    \mathcal{U} = \big\{ \mathbf{u} \in \reals^{6 \times 1} \bigr| | \mathbf{u}_k^\top \cdot \mathbf{r}_\alpha \times d_g| \leq \bar{u} \big\},    
\end{align}
where $\bar{u} \in \reals_+$ is the upper limits of the control input from the manipulator for all time $k \in \reals_+$.
This constraint limits the force applied to the hinge to avoid damaging the jammed component. 
Enforcing this constraint is challenging since the contact point $d_g$ is unknown to the controller.  

Likewise, to prevent damage, we need to enforce the state constraints 
\begin{align}\label{eq:state_constraint}
    \mathcal{X} = \big\{ \mathbf{x} \in \reals^2 \bigr|  \mathbf{x}_l \leq \mathbf{x}_k \leq \mathbf{x}_u \big\}.
\end{align}
\end{subequations}
where $\mathbf{x}_l = [\theta_l, \dot{\theta}_l]^\top \in \reals^{2 \times 1}$ and $\mathbf{x}_u = [\theta_u, \dot{\theta}_u]^\top \in \reals^{2 \times 1}$ denote the upper bound and lower bound of the state vector for all time $k \in \reals_+$, respectively. The bounds $\mathbf{x}_l$ and $\mathbf{x}_u$ prevent overextending the hinge or causing the solar panel from colliding with another part of the client or servicer. 
Enforcing this constraint is challenging, since preventing over-extension may require large interaction forces that could overstress the \textit{Client} structure.



\subsection{Safe On-Orbit Dislodging Problem}
The safe on-orbit dislodging problem can be described as the servicer dislodges the client to regulate the client state $\mathbf{x}_k$ from the initial condition $\mathbf{x}_0$ to the desired equilibrium $\mathbf{x}^*$. Meanwhile, the parameter estimator can estimate the unknown time-varying parameter $\boldsymbol{\rho}_k$ to the true value $\boldsymbol{\rho}^*$ during the manipulation process. The objective can be described as  
\begin{equation}
\begin{aligned}
\label{eq::assump_goal}
\left\{
\begin{array}{l}
 \mathbf{e}_x = \displaystyle\lim_{k \rightarrow \infty} \norm{\mathbf{x}^*{-}\mathbf{x}_k}_2,\\
 \mathbf{e}_\rho = \displaystyle\lim_{k \rightarrow \infty} \: \norm{\boldsymbol{\rho}^*{-}\boldsymbol{\rho}_k}_2.
\end{array}
\right.
\end{aligned}
\end{equation}
where $\mathbf{e}_x \in \reals_+$ and $\mathbf{e}_\rho \in \reals_+$ denote the state and parameter estimate error, respectively. While enforcing the safety constraint  ~\eqref{eq:ParameterBounds},~\eqref{eq::disturbance_constraint},~\eqref{eq:Constraints}, the hinge state should converge to the desired equilibrium and the uncertainty parameters should converge to their actual values. This requires balancing between the exploration to learn the uncertain parameters with bounded estimation errors and exploitation to use these bounds to ensure robust safety during dislodging.

\section{Robust Adaptive MPC for Dislodging}
In this section, we introduce our robust adaptive MPC algorithm. This includes its set-membership parameter estimator,  robust tube constraints, robust terminal set, and dual-mode cost function.

\subsection{Linear Parametric Modeling}
To facilitate learning the uncertainty parameters $\boldsymbol{\rho}_k$ and build up the connection for the solar panel between the rotation along piovt axis relative to frame $\mathcal{L}$ in 1 DoF and the motion relative to the frame $\mathcal{S}$ in 6 DoF,  we reorganize the hinge dynamics~\eqref{eq::Dynamics_euler} into the linear parametric state-space 
\begin{equation} \label{eq:parametric-plant}
    \mathbf{x}_{k+1}= \left( \mathbf{I} + h \mathbf{A}(\boldsymbol{\rho}_k) \right) \mathbf{x}_k+h\mathbf{B}(\boldsymbol{\rho}_k) \mathbf{u}_k+\begin{bmatrix}
0 \\
{\dfrac{-\tau_f}{M}}
\end{bmatrix} + \mathbf{w}_k,
\end{equation}
where  $\mathbf{I} \in \reals^{2 \times 2}$ is the identity matrix, $\mathbf{x}_k = [\theta_k, \dot{\theta}_k]^\top {\in} \reals^2$ and $\mathbf{w}_k = \left[0, \tfrac{\tau_{m_k}}{M}\right]^\top {\in} \reals^2$ denotes the disturbance which is bounded. $h \in (0, 1)$ is the sampling period for discretization of the state matrices $\mathbf{A}(\boldsymbol{\rho}_k)$ and $\mathbf{B}(\boldsymbol{\rho}_k)$, which can be parameterized as 
\begin{equation} \label{eq:parm}
\begin{aligned}
            \mathbf{A}(\boldsymbol{\rho}_k) &= \mathbf{A}_0 + \displaystyle\sum_{i=1}^{n} \mathbf{A}_i [\boldsymbol{\rho}_k]_i,\\ \quad  \mathbf{B}(\boldsymbol{\rho}_k) &= \mathbf{B}_0 + \displaystyle\sum_{i=1}^{n} \mathbf{B}_i [\boldsymbol{\rho}_k]_i.
            \end{aligned}
\end{equation}
where $n = 3$ denotes the number of unvertain parameter in $\boldsymbol{\rho}_k$. Then based on~\eqref{eq::Dynamics_euler}, we can get~\eqref{eq:parm} in detail as 

\begin{equation*}
\begin{aligned}
\mathbf{A}(\boldsymbol{\rho}_k) &= \begin{bmatrix}
0 & 1 \\
-\dfrac{\mathcal{K}_k}{M} & -\dfrac{\zeta_k}{M}
\end{bmatrix} \\
    \mathbf{B}(\boldsymbol{\rho}_k) &=  \frac{1}{M}\left[\begin{array}{cccccc}
0 & 0 & 0 & 0 & 0 & 0 \\
-\mathcal{D}_k & \mathcal{L}_k & 0 & 0 & 0 & 1
\end{array}\right] 
\end{aligned}
\end{equation*}
Note that considering the gripper may not be perpendicular to the rod for all time during the dislodging task, $\alpha$ is also time-varying. Here, we choose $\phi$ to denote the deviation of the EE's angle from its ideal position which is perpendicular to the central axis of the solar panel. Also, we assume the $\alpha$ would not deviate more than $70^\circ$ from the initial position so that $\phi = 90^\circ - \alpha \leq 20^\circ$. Under this assumption, based on $\sin \theta=\sin (\pi+\phi) \approx-\phi$ and $\cos \theta=\cos (\pi+\phi) \approx-1$, we can use $-\phi$ and $-1$ to replace $ \sin(\alpha_k)$ and $\cos(\alpha_k)$ respectively to get $\mathcal{D}_k = -d_g \cdot \phi_k$ and $\mathcal{L}_k = -1$. 

\subsection{Set-Based Parameter Estimation}

In this section, we present our set-based parameter estimator for dislodging which bounds the uncertain parameters $\boldsymbol{\rho}_k$ of the linear parametric model~\eqref{eq:parametric-plant} using real-time state $\mathbf{x}_k$ of the client and control input $\mathbf{u}_k$ from the servicer. Our robust adaptive MPC will leverage these bounds to ensure robust constraint satisfaction despite parametric uncertainty. To bound the parameter uncertainty, we make the following assumption about the boundedness of the measurement noise $\mathbf{w}_k$. 

\begin{assumption}
\label{asmp::1}
The disturbance set $\mathbb{W}$ is a bounded polytope which
includes exogenous disturbances and small unmodeled actuator nonlinearities, e.g., deadzones, backlash, etc, at the end-effector or hybrid hinge as described by the $ n_w $ constraints in the set 
\begin{equation}
    \mathbb{W} = \{\mathbf{w}_k \in \mathbb{R}^2 | \mathbf{H}_w \mathbf{w}_k \le \boldsymbol{\delta}_w \}.
    \label{eq::disturbance_constraint}
\end{equation}
\end{assumption}
where $\mathbf{H}_w \in \reals^{n_w \times 2}$ and $\boldsymbol{\delta}_w \in \reals^{n_w}$.
Note that we assume that no deadzone nonlinearity is modeled explicitly, and its effect is treated as an additive bounded disturbance within $\mathbb{W}$.
The parameter set $\boldsymbol{\rho}_k$ is iteratively updated at each time step $k$ to form a set $\mathcal{P}_k$ which bounds all possible values of the uncertain parameters $\boldsymbol{\rho}_k$. This update is achieved by constructing a non-falsified parameter set, utilizing measurement data from the preceding $s$ time steps $\nu_s$, as follows:

\begin{equation} \label{eq:SimpleNonfalsified}
\begin{aligned}
\Delta_{k} &:= \biggr\{
\boldsymbol{\rho}_k \in \mathbb{R}^{p}\: \biggr|\mathbf{x}_{\nu_s+1} -\left(\mathbf{A}(\boldsymbol{\rho}_k) \mathbf{x}_{\nu_s} + \mathbf{B}(\boldsymbol{\rho}_k) \mathbf{u}_{\nu_s} \right) \in \mathbb{W}
\biggr\}\\ 
&=\biggr\{
\boldsymbol{\rho}_k \in \mathbb{R}^{p}\: \biggr| \: -\mathbf{H}_w \mathbf{D}_{{\nu}_s{-}{1}} \boldsymbol{\rho}_k  \le \boldsymbol{\delta}_w+\mathbf{H}_w \mathbf{d}_{k}\biggr\}, 
\end{aligned}
\end{equation}
where $\forall \nu_s {\in} \{k{-}s,k{-}1\}_{\mathbb{N}}$ denotes the preceding time steps from time $k{-}s$ to $k{-}1$, and $\mathbf{D}_{\nu_s} {\in} \mathbb{R}^{n\times 3}$ can be defined as 
\setlength\arraycolsep{2pt}
\begin{equation}\label{eq:Dtdt}
\resizebox{0.87\hsize}{!}{$
    \mathbf{D}_{\nu_s}(\mathbf{x}_{\nu_s},\mathbf{u}_{\nu_s}) = \begin{bmatrix}
         \mathbf{A}_1 \mathbf{x}_{\nu_s} + \mathbf{B}_1 \mathbf{u}_{\nu_s}, & \ldots, & \mathbf{A}_3 \mathbf{x}_{\nu_s} + \mathbf{B}_3 \mathbf{u}_{\nu_s}
    \end{bmatrix}.$}
\end{equation}
and $\mathbf{d}_{{\nu_s}+1} = \mathbf{A}_0 \mathbf{x}_{\nu_s} + \mathbf{B}_0 \mathbf{u}_{\nu_s} - \mathbf{x}_{{\nu_s}{+}{1}}$. Note that $\mathbf{D}_{\nu_s}$ and $\mathbf{d}_{\nu_s}$ are quantities that linearly depend on the measured state and input vectors, but the dependence is omitted for clarity. In this notion case, we can represent $\Delta$ using hyperplane constraints in $ \mathbb{R}^n $, i.e., $\Delta$ is polytopic.

In \eqref{eq:SimpleNonfalsified}, the non-falsified set $\Delta_{k}$ is constructed by performing a consistency check for every candidate parameter against the latest measurement data.  This is achieved by using the measured sequence $ \{\mathbf{x}_{k-s},\ldots,\mathbf{x}_{k}\}$ and inputs to calculate the disturbance $\mathbf{w}_k$  that a candidate parameter $\boldsymbol{\rho}_k$ would imply. The parameter is then 'falsified' if this disturbance falls outside the known bounded set $\mathbb{W}$. This process results in a set of linear inequalities that geometrically define $\Delta_k$ as a computationally tractable polytope, representing all parameters consistent with the observed system behavior.  
To manage computational complexity, the polytopic set $\mathcal{P}_{k}$ is defined using a fixed number of linear constraints 
\begin{equation} \label{eq:Theta_k_def}
\mathcal{P}_k :=  \{\boldsymbol{\rho}_k \in \mathbb{R}^3| \mathbf{H}_{\rho_k} \boldsymbol{\rho}_k\le \boldsymbol{\delta}_{\rho_k}\},
\end{equation}
where the fixed matrix $\mathbf{H}_{\rho_k} \in \reals^{n_p \times 3}$ is chosen offline and $ \boldsymbol{\delta}_{\rho_k} \in \reals^{n_p}$ is updated online. 
Note that as new measurements become available, the number of constraints defining the parameter set would grow indefinitely, making the problem computationally intractable for a real-time controller. To manage this computational complexity, the polytopic set $\mathcal{P}_k$ in~\eqref{eq:Theta_k_def} using a fixed number of linear constraints, where the matrix $\mathbf{H}_{\rho_k}$ chosen offline. Rather than adding new constraints with each measurement, the algorithm simply updates the vector $\boldsymbol{\delta}_{\rho_k}$ online, which tightens this fixed-structure polytope based on new information. To account for the time-varying~\eqref{eq:parameter-velocity} parameters $\boldsymbol{\rho}_k^*$, we introduce a dilation operator with $\varsigma \in \mathbb{N}$ for $\mathcal{P}$ as
\begin{equation}
    \mathbf{d}_{\varsigma}(\mathcal{P}) :=\left\{\boldsymbol{\rho}_k \in \mathbb{R}^p \mid \mathbf{H}_{\rho_k} \boldsymbol{\rho}_k \leq \boldsymbol{\delta}_\rho+ \varsigma \boldsymbol{\varphi}_\rho \boldsymbol{\gamma}\right\},
    \label{eq::dia_vec}
\end{equation}
where the vector $\boldsymbol{\gamma} \in \reals_+^{n_\gamma}$ and which dilates the constraints by a factor of $\varsigma \in \mathbb{N}$ in the direction $\boldsymbol{\varphi}_\rho$. 
Using the dilation operator~\eqref{eq::dia_vec}, we have following the update rule for the uncertainty sets $\mathcal{P}_k$ bounding the parameter $\boldsymbol{\rho}_k$ 
\begin{equation}
\label{eq:: para_update_rule}
    \mathcal{P}_{k+1} := \mathbf{d}_{\varsigma}\left(\mathcal{P}_{k}  \cap \Delta_{k+1}\right) \cap \mathcal{P},
\end{equation}
 at time step $k$. This dialation operator robustly inflates the current parameter set $\mathcal{P}_{k}$ to account for potential changes, ensuring the true parameter at the next step $\boldsymbol{\rho}_{k+1}^*$ remains contained within the set. The size of this inflation is governed by the maximum parameter drift rate $\varphi_\rho$ from~\eqref{eq:parameter-velocity}  making this a critical step for providing robust safety guarantees. 
Our robust adaptive MPC requires that we predict how the bounds $\hat{\mathcal{P}}_{\varsigma|k}$ on the time-varying parameters $\boldsymbol{\rho}_k$ evolve over the prediction horizon. These prediction bounds are given by
\begin{equation}
\label{eq::para_est_dia}
    \hat{\mathcal{P}}_{\varsigma \mid k}=\mathbf{d}_{\varsigma}\left(\mathcal{P}_{_k}\right) \cap \mathcal{P},
\end{equation}
for $\varsigma = 1,\dots,N-1$ where $\hat{\mathcal{P}}_{\varsigma \mid k}$ is the predicted uncertainty bound at time $k+\varsigma$. Note that the recursive parameter update in~\eqref{eq::para_est_dia} is designed to accommodate initial parameter uncertainty. The set-membership identification scheme actively refines the parameter estimate by iteratively tightening the admissible parameter set through newly acquired measurements. This inherent characteristic ensures that the robust safety of the system is guaranteed and performance is maintained even when commencing from a coarse initial parameter set that does not heavily rely on the precise initial set chosen.

This is ensured by calculating $\boldsymbol{\delta}_{\rho_k} $ as a solution to the following set of linear programs:
\begin{equation}
\begin{aligned}
\begin{split}
[\boldsymbol{\delta}_{\rho_k}]_i \: &=\: \min_{\boldsymbol{\rho}\in \mathbb{R}^{p}} \quad  [\boldsymbol{\delta}_{\rho}]_i\boldsymbol{\gamma}_{i}, \quad \forall i \in \{1,n_{\rho^+}\}_{\mathbb{N}}
 \label{eq:Theta_k_LP}\\
 &\text{s. t. } \quad  
 \begin{bmatrix}
    \mathbf{H}_\rho \\ -\mathbf{H}_w \mathbf{D}_{\nu_s}
\end{bmatrix} \boldsymbol{\rho}^+
 \le \begin{bmatrix}
  \boldsymbol{\delta}_{\rho_{k-1}} \\ \boldsymbol{\gamma}+\mathbf{H}_w \boldsymbol{d}_{k+1}
 \end{bmatrix}. 
\end{split}
\end{aligned}
\end{equation}
Note that Eq.~\eqref{eq:Theta_k_LP} is an outer-approximation of Eq.~\eqref{eq:SimpleNonfalsified} intended to reduce computational complexity.
\subsection{Robust Tube Constraints}

In this section, we derive a state-space tube $\mathbb{X}_{\varsigma|k}$ for $\varsigma$ 
and we can guarantee contains the state $\mathbf{x}_k$ despite parametric uncertainty $\boldsymbol{\rho}_k \in \mathcal{P}_k$ and disturbances $\mathbf{w}_k \in \mathbb{W}$. By remaining in the state tube, we can guarantee that the servicer would not damage the client. The states of the client and the control inputs from the servicer must satisfy the constraints as
\begin{equation} \label{eq:Constraints}
\begin{aligned}
\mathbb{Z} &= \left\{(\mathbf{x}_k,\mathbf{u}_k) \in \mathbb{R}^n \times \mathbb{R}^m \bigr|  \mathbf{F} \mathbf{x}_k + \mathbf{G} \mathbf{u}_k \le \boldsymbol{\gamma}\right\},\\
\end{aligned}
\end{equation} 
where the matrices $\mathbf{F} \in \mathbb{R}^{n_\gamma \times n}$ and $\mathbf{G} \in \mathbb{R}^{n_\gamma \times n}$ are derived from the input~\eqref{eq:input_constraint} and state~\eqref{eq:state_constraint} constraints. The objective is to regulate the \textit{Client} state $\mathbf{x}_k \in \reals^2$ from the initial condition $\mathbf{x}_0$ to the desired state $\mathbf{x}_d$, and the constraints the control input $\mathbf{u}_k \in \reals^6$ from the \textit{Servicer} under safety set,  while robustly satisfying the constraints in \eqref{eq:Constraints}. 

The tube MPC approach proposed in~\cite{lopez2019dynamic} ensures robust constraint satisfaction. The control input is parameterized using a feedback gain $\mathbf{K} \in \reals^{m \times n}$ as
\begin{equation}
\mathbf{u}_{\varsigma|k} = \mathbf{K} \mathbf{x}_{\varsigma|k} + \mathbf{v}_{\varsigma|k},
\label{eq:tubeMPC_1}
\end{equation}
where $\mathbf{v}_{N \mid k}=\left\{\mathbf{v}_{\varsigma \mid k}\right\}_{j \in \{0,N-1\}_\mathbb{N}} {\in} \reals^m$ are decision variables in the MPC optimization problem. We make the following standard assumption about the feedback gain $\mathbf{K}$. 
\begin{assumption}
\label{asmp::2}
The feedback gain $\mathbf{K}$ is chosen such that $ \mathbf{A}_\text{cl}(\boldsymbol{\rho}_k)  = \mathbf{A}(\boldsymbol{\rho}_k) + \mathbf{B}(\boldsymbol{\rho}_k)\mathbf{K}$ is asymptotically stable $ \forall \boldsymbol{\rho}_k \in \mathcal{P} \supseteq \mathcal{P}_k$.
\label{asp::feedback}
\end{assumption}

A gain $\mathbf{K}$ satisfying Assumption~\ref{asmp::2} can be computed using standard robust control techniques, e.g., following the the linear matrix inequality from~\cite{wang1992robust}.

The state tube is defined using the set-based dynamics
\begin{subequations}\label{eq:SetDynamics}
\begin{align}
\mathbf{x}_{k} &\in \mathbb{X}_{0|k},  \label{eq:initialSet}\\
\mathbf{A}_\text{cl}(\boldsymbol{\rho}_k)\mathbb{X}_{\varsigma|k}  \oplus \mathbf{B}(\boldsymbol{\rho}_k)\mathbf{u}_{\varsigma|k} \oplus \mathbb{W} &\subseteq \mathbb{X}_{\varsigma+1|k} , \label{eq:SetInclusions} 
\end{align}
\end{subequations}
for $\varsigma=0,\dots,N-1$ and for all parameters $\boldsymbol{\rho}_k \in \mathcal{P}_{\varsigma|k}$. 
This ensures that $\mathbf{x}_{\varsigma|k} \in \mathbb{X}_{\varsigma|k} $ for all the realizations of uncertainty and disturbance. To manage computational complexity, the tube cross-section at each time step, $\left\{\mathbb{X}_{\varsigma \mid k}\right\}_{j \in \{0,N\}_\mathbb{N}}$, is parameterized by translating and scaling of the set 
\begin{align}
\label{eq:cross-section}
 \mathbb{X}_0^\curlyvee := \{\mathbf{x} \in \reals^n | \mathbf{H}_x^\curlyvee \mathbf{x} \le \boldsymbol{\gamma}\},
\end{align}
where the fixed matrix $\mathbf{H}_x$ is selected offline. 
Then, for $\varsigma = \{0,N\}_\mathbb{N}$, the state tube is parameterized as
\begin{equation}\label{eq:StateTubeParameterization}\resizebox{0.87\hsize}{!}{$
\begin{aligned}
\mathbb{X}_{\varsigma|k} &= \{\mathbf{z}_{\varsigma|k}\} \oplus \mathbb{X}_{0,k}^\curlyvee = \{\mathbf{x} \in \reals^n| \mathbf{H}_{x,k}^\curlyvee (\mathbf{x}-\mathbf{z}_{\varsigma|k}) \le \boldsymbol{\gamma}\}\\
&= \{\mathbf{z}_{\varsigma|k}\} \oplus \vartheta_{\varsigma|k} \text{co}\{\mathbf{x}^{1},\mathbf{x}^{2},\ldots,\mathbf{x}^{v}\}.
\end{aligned}$}
\end{equation}
where the $\{\mathbf{x}^{1},\mathbf{x}^{2},\ldots,\mathbf{x}^{v}\}$ are the vertices of the polytope~\eqref{eq:cross-section}. The variables $\mathbf{z}_{\varsigma|k} \in \mathbb{R}^{n}$ and $\vartheta_{\varsigma|k} \in \mathbb{R}_{+}$ are decision variables in the MPC optimization, which
respectively define the translation and scaling of $\mathbb{X}_{0}$. 

\subsection{Reformulation of Robust Tube Constraints}\label{SetDynamics}

\begin{figure}[!t]
\centering
\captionsetup{font=footnotesize}
\includegraphics[width=0.45\textwidth]{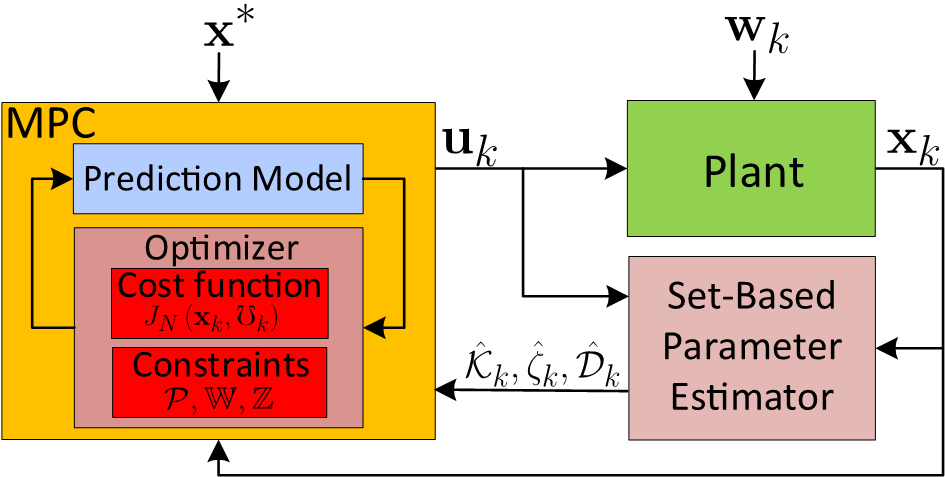}
\caption{The block diagram of our robust adaptive MPC algorithm.}
\label{fig::block_diagram_mpc}
\end{figure}

In this section, we present a convex formulation of the robust tube $\mathbb{X}_{\varsigma|k}$ which can be integrated into the MPC constraints to produce a tractable optimization problem. 
The state and input constraints defined in \eqref{eq:Constraints} and the set dynamics proposed in \eqref{eq:SetDynamics} must be robustly satisfied for all $\boldsymbol{\rho} \in \mathcal{P}_k$ and disturbances $\mathbf{w} \in \mathbb{W}$. To reformulate these in a convex manner, the following notation is defined
\begin{subequations}\label{eq:x_jlk}
\begin{align}
\mathbf{d}_{\varsigma|k}^{i} &= \mathbf{A}_0 \mathbf{x}_{\varsigma|k}^{i} + \mathbf{B}_0 \mathbf{u}_{\varsigma|k}^{i} - \mathbf{z}_{\varsigma+1|k},\\ \mathbf{D}_{\varsigma|k}^{i} &= \mathbf{D}(\mathbf{x}_{\varsigma|k}^{i},\mathbf{u}_{\varsigma|k}^{i}),
\end{align}
\end{subequations}
where $\mathbf{x}_{\varsigma|k}^{i} = \mathbf{z}_{\varsigma|k} + \vartheta_{\varsigma|k} \mathbf{x}^{i}$, $\mathbf{u}_{\varsigma|k}^{i} = \mathbf{K}\mathbf{x}_{\varsigma|k}^{i} + \boldsymbol{c}_{\varsigma|k}$, $i \in\{1,v\}_\mathbb{N} ,\varsigma \in \{0,N{-}1\}_\mathbb{N}$. Note that unlike the definition in \eqref{eq:Dtdt} where $\mathbf{D}_{\nu_s},\mathbf{d}_{\nu_s}$ are a function of known states and inputs, the quantities $\mathbf{D}_{\varsigma|k}^{j}, \mathbf{d}_{\varsigma|k}^{j}$ are linearly depend on the decision variables of MPC. Additionally, we define the vectors $ \bar{\mathbf{f}}$ and $\bar{\mathbf{w}} $, which are computed offline such that for $i \in \{1,n_c\}_\mathbb{N},j \in \{1,n_x\}_\mathbb{N}$
\begin{equation}
\begin{aligned}
 [\bar{\mathbf{f}}]_{i} &= \displaystyle\max_{x\in \mathbb{X}_0} [\mathbf{F}+\mathbf{G}\mathbf{K}]_i \mathbf{x},\\ [\bar{\mathbf{w}}]_{j} &= \displaystyle\max_{w\in \mathbb{W}} \: [\mathbf{H}_x]_j \mathbf{w}.
\end{aligned}
\end{equation}
From~\cite{kouvaritakis2016model},  we can reformulate the robust tube constraints~\eqref{eq:StateTubeParameterization} as linear equality and inequality constraints as follows. Let the state tube $ \{\mathbb{X}_{\varsigma|k}\}_{\varsigma \in \{{0},{N}\}_{\mathbb{N}}} $ be parameterized according to \eqref{eq:StateTubeParameterization}. Then, the constraints \eqref{eq:Constraints} and set-dynamics \eqref{eq:SetDynamics} are satisfied if and only if $\forall  j \in \{1,v\}_\mathbb{N}$, $ \varsigma \in \{0,N{-}1\}_\mathbb{N}$ there exists $\boldsymbol{\Gamma}_{\varsigma|k}^{j} \in \mathbb{R}^{n_x\times n_p}_{+}$ such that 
	\begin{subequations}\label{eq:lambdaConstraints}
		\begin{align}
		(\mathbf{F}+\mathbf{G}\mathbf{K})\mathbf{z}_{\varsigma|k} + \mathbf{G}\mathbf{v}_{\varsigma|k} + \vartheta_{\varsigma|k}\bar{\boldsymbol{f}} &\le \boldsymbol{\gamma},\\
		-\mathbf{H}_x \mathbf{z}_{0|k} -\vartheta_{0|k}\boldsymbol{\gamma} &\le -\mathbf{H}_x \mathbf{x}_k ,		\label{eq:x0Constraint}\\
		\boldsymbol{\Gamma}_{\varsigma|k}^{j} \boldsymbol{\delta}_{\rho_k} + \mathbf{H}_x \mathbf{d}_{\varsigma|k}^{j} -\vartheta_{\varsigma+1|k} \boldsymbol{\gamma} &\le -\bar{\mathbf{w}} ,\label{eq:InclusionIneq}\\
		\mathbf{H}_x \mathbf{D}_{\varsigma|k}^{j} &= \boldsymbol{\Gamma}_{\varsigma|k}^{j} \mathbf{H}_{\rho_k} \label{eq:InclusionEqual}.
		\end{align} 
	\end{subequations}
These linear inequality constraints allow the state tube constraints~\eqref{eq:StateTubeParameterization} to be incorporated as tractable constraints in the MPC optimization.

Similarly, we can also get the predicted state tube $ \{\hat{\mathbb{X}}_{\varsigma|k}\}_{\varsigma \in \{{0},{\hat{N}-1}\}_\mathbb{N}}$ satisfy the set-dynamics \eqref{eq:SetDynamics} if and only if for all $j \in  \{1,v\}_\mathbb{N}$ and $\varsigma \in  \{0,\hat{N}{-}1\}_\mathbb{N}$ there exists $ \hat{\boldsymbol{\Gamma}}_{\varsigma|k}^{i} \in \mathbb{R}^{n_x\times (n_p+n_w)}_{+}$ such that 
\begin{subequations}\label{eq:lambda_tilConstraints}
\begin{align}
-\mathbf{H}_x \hat{\mathbf{z}}_{0|k} - \hat{\vartheta}_{0|k}\boldsymbol{\gamma} &\le -\mathbf{H}_x \mathbf{x}_k, \\
\hat{\mathbf{K}}_{\varsigma|k}^{i} \hat{\boldsymbol{\delta}}_{\rho_k} + \mathbf{H}_x \hat{\mathbf{d}}_{\varsigma|k}^{i} -\hat{\vartheta}_{\varsigma+1|k} \boldsymbol{\gamma} &\le -\bar{\mathbf{w}},\\
\mathbf{H}_x \hat{\mathbf{D}}_{\varsigma|k}^{i} &= \hat{\boldsymbol{\Gamma}}_{\varsigma|k}^{i} \hat{\mathbf{H}}_{\rho_{\varsigma|k}}.
\end{align}
\end{subequations} 
where $\hat{\mathbf{H}}_{\rho|k}=\left[\mathbf{\hat H}_{\rho_k} \  \mathbf{\hat H}_\rho\right] \in \mathbb{R}^{q_k^{\prime} \times p}$ and $\hat{\boldsymbol{\delta}}_{\rho_k}=\left[\boldsymbol{\delta}_{\rho_k}+ \varsigma \boldsymbol{\gamma} \  \boldsymbol{\delta}_\rho \right] \in \mathbb{R}^{n_p}$. The constraints are bilinear in the variables since $ \hat{\mathbf{H}}_{\rho}, \hat{\boldsymbol{\delta}}_{\rho_k} $ are linearly dependent on the control input $\mathbf{u}_k $ as seen in \eqref{eq:Theta_k_def}.



\subsection{Robust Terminal Set}

In this section, we derive a terminal set $\mathbb{X}_\text{T}$ which we will use to ensure our robust adaptive MPC is recursively feasible despite parametric uncertainty $\boldsymbol{\rho}_k \in \mathcal{P}$ and disturbances $\mathbf{w}_k \in \mathbb{W}$. 
Terminal constraints are imposed on $\mathbf{z}_{N|k}$ and $\vartheta_{N|k}$ so that the state tube constraints are directed into the terminal set to ensure the dislodging process in its final step does not lead the system into a region from which it could become unsafe in the future. We make the following assumption about the existence of a terminal set. 
\begin{assumption}
\label{asmp::3}
	There exists a nonempty terminal set $  \mathbb{X}_{T} = \{(\mathbf{z},\vartheta)\in \mathbb{R}^n{\times}\mathbb{R} | \: \mathbf{z} {=} 0,\: \vartheta {\in} [0,\bar{\vartheta}] \}$, such that  for all $ \boldsymbol{\rho}\in \mathcal{P} $ it holds that
\begin{align*}
(\mathbf{z}, \vartheta) \in \mathbb{X}_\text{T} \Rightarrow & \exists\left(\mathbf{z}^{+}, \vartheta^{+}\right) \in \lambda \mathbb{X}_\text{T} \text { s.t. } \\ & \mathbf{A}_\text{cl}(\boldsymbol{\rho})\left(\{\mathbf{z}\} \oplus \vartheta \mathbb{X}_0\right) \subseteq\left\{\mathbf{z}^{+}\right\} \oplus \vartheta^{+} \mathbb{X}_0 \\ (\mathbf{z}, \vartheta) \in \mathbb{X}_\text{T} \Rightarrow & \exists\left(\mathbf{z}^{+}, \vartheta^{+}\right) \in \mathbb{X}_\text{T} \text { s.t. } \\ & \mathbf{A}_\text{cl}(\boldsymbol{\rho})\left(\{\mathbf{z}\} \oplus \vartheta \mathbb{X}_0\right) \oplus \mathbb{W} \subseteq\left\{\mathbf{z}^{+}\right\} \oplus \vartheta^{+} \mathbb{X}_0 \\ (\mathbf{z}, \vartheta) \in \mathbb{X}_\text{T} \Rightarrow & (\mathbf{x}, \mathbf{K}\mathbf{x}) \in \mathbb{Z} \quad \forall \mathbf{x} \in\{\mathbf{z}\} \oplus \vartheta \mathbb{X}_0
\end{align*}
\end{assumption}

Assumption \ref{asmp::3} implies that the set $ \mathbb{X}_{T} $ is a robust positively invariant (RPI) set~\cite{kolmanovsky1998theory} for the set-dynamics in $ (\mathbf{z},\vartheta) $, with an additional constraint that the set $ \mathbb{X}_{N|k} $ remains centered at origin which is the desired equilibrium point $\mathbf{x}^*$ for dislodging problem. Note that Assumption \ref{asp::feedback} is a necessary condition for Assumption \ref{asmp::3} to be satisfied, but they are stated separately to emphasize that the stronger assumption is only needed to implement the terminal condition.


\subsection{Robust Adaptive MPC Algorithm}

In this section, we present our robust adaptive MPC algorithm for the safe on-orbit dislodging problem.

First, we define our double-safe cost function formulated as
\begin{equation}
J_N(\mathbf{x}_{k},\boldsymbol{\mho}_k)= \sum_{i=0}^{N-1} \ell\left(\mathbb{X}_{i \mid k}, \mathbf{v}_{i \mid k}\right)+ \ell\left(\mathbb{X}_{N \mid k}\right),
\label{eq:MPCcost}
\end{equation}
with $\boldsymbol{\mho} = \bigl \{\{\mathbf{z}_{\varsigma|k}\},\{\vartheta_{\varsigma|k}\},\{\mathbf{v}_{\varsigma|k}\},\{\mathbf{K}^{j}_{\varsigma|k}\} \bigl \}$ as the decision variable. $\ell\left(\mathbb{X}_{i \mid k}, \mathbf{v}_{i \mid k}\right) = \max_{\mathbf{x}\in\mathbb{X}} ||\mathbf{Q}\mathbf{x}||_{\infty} + ||\mathbf{R}(\mathbf{K}\mathbf{x}+\mathbf{v})||_{\infty} + ||\boldsymbol{\Upsilon}\boldsymbol{\rho}||_{\infty}$, and $ \mathbf{Q},\mathbf{R},\boldsymbol{\Upsilon}$ are positive definite matrices. Note that in this work, the term introduced $||\boldsymbol{\Upsilon}\boldsymbol{\rho}||_{\infty}$ explicitly optimizes not only for the system's control performance but also for improving the accuracy of your time-varying parameter estimates. This additional objective term penalizes the size of the predicted parameter uncertainty set. The optimizer is thereby guided to select informative control actions that enhance the persistency of excitation and accelerate the convergence of the parameter bounds. The framework consequently computes a single control policy, inherently balancing the exploitative task of achieving control performance with the exploratory task of improving parameter identification. The results for parameter estimation in Section~\ref{sec::sim_results} evidently shows the advantage of this novel contribution compared with traditional adaptive MPC.    
A linear cost function is selected to enable its reformulation through linear inequalities resulting in the MPC optimization problem becoming a linear program. 



\begin{algorithm} [!b]
    \caption{Robust Adaptive MPC in parameter exploration}
    \begin{algorithmic}[1]
        \Statex \textbf{Input} Choose $\mathbf{K}$, $\bar{\vartheta}$, and $\mathbb{X}_0$. Initialize $\boldsymbol{\delta}_{\rho_k}$ and $\hat{\boldsymbol{\rho}}_k$. Determine $ \mathbf{Q},\mathbf{R},\boldsymbol{\Upsilon}$ for~\eqref{eq:MPCcost}
        \Statex \textbf{Online}
        \State $k \gets 0$
        \For{$k = 0$ to $N$}
            \State Obtain the measurement $\mathbf{x}_k$
            \State Construct $\Delta_k$ according to \eqref{eq:SimpleNonfalsified}
            \State Update $ \mathcal{P}_k$ using \eqref{eq:: para_update_rule}
            \State Update $\boldsymbol{\delta}_{\rho_k}$ using \eqref{eq:Theta_k_LP} and compute $\hat{\boldsymbol{\rho}}_k$ 
            \State Solve optimization problem \eqref{eq::optimazation_min}
            \State Apply the control input $\mathbf{u}_k = \mathbf{K} \mathbf{x}_k + \mathbf{v}_{0|k}$
            \State $k \gets k + 1$
        \EndFor
    \end{algorithmic}
\label{psudo::rampc_parameter}
\end{algorithm}
Then, the optimization problem can be written as

\begin{equation}
\begin{array}{r}
 \min J_N\left(\mathbf{x}_{k},\boldsymbol{\mho}_k \right) \\
\text { s.t. } \eqref{eq:lambdaConstraints},\eqref{eq:lambda_tilConstraints} .
\end{array}
\label{eq::optimazation_min}
\end{equation}
The block diagram of the whole algorithm as shown in Fig.~\ref{fig::block_diagram_mpc} and the following proposition show that our robust adaptive MPC enforces the dislodging constraints despite parametric uncertainty.

\begin{table*}[!b]
    \centering
    \footnotesize
    \setlength{\tabcolsep}{5pt}  
    \captionsetup{font=footnotesize}
    \begin{tabular}{ccccccccccc}
        \toprule
        Parameter & $m$ & $d_s$ & $d_g$ & $\mu$ & $r_e$ & $I_{cm}$ & $J_{m}$ & $\mathcal{K}_k$ & $\zeta_k$ & $\mathcal{D}_k$\\
        \midrule
        Value & 1.8\,kg & 1.2\,m & 2.5\,m & 0.3 & 0.2\,m & 1.5\,kg/m$^2$ & 4.1\,kg/m$^2$ & $0.45 + 0.1 \sin(0.1k)$ & $0.6 + 0.15 \sin(0.1k)$ & $2.5 + 0.2 \sin(0.1k)$\\
        \bottomrule
    \end{tabular}
    \caption{Simulation parameters for the time-varying system in the dislodging task.}
    \label{tab:rampc_para}
\end{table*}

\textit{Theorem 1:}
\label{prop::close-loop}
Suppose the \textit{Assumptions} \ref{asmp::1}, \ref{asmp::2} and \ref{asmp::3} be satisfied and an initial feasible solution exist. 
Let the parameter estimate set $\mathcal{P}_k$ be defined by the set dynamics \eqref{eq:: para_update_rule} and  admissible input set $\mathcal{I}_{v_s}\left(\mathbf{x}_k, \mathcal{P}_k\right)$ as
\begin{equation}
\resizebox{0.85\hsize}{!}{$
\mathcal{I}_{v_s}\left(\mathbf{x}_k, \mathcal{P}_k\right)=\left\{\boldsymbol{\mho}_k \mid \text { \eqref{eq:lambdaConstraints}, \eqref{eq:lambda_tilConstraints}, }\left(z_{N \mid k}, \vartheta_{N \mid k}\right) \in \mathbb{X}_\text{T}\right\}$}
\end{equation}
Then, the system~\eqref{eq:parametric-plant} with the proposed MPC control law in~\eqref{eq:tubeMPC_1} using Algorithm \ref{psudo::rampc_parameter} satisfies the following closed-loop properties for all $ k > k_0 $:
\begin{enumerate}
	\item $ \boldsymbol{\rho}^*_k \in \mathcal{P}_k$
	\item $ \mathcal{I}_{v_s} \neq \{\varnothing\} $
	\item $ (\mathbf{x}_k,\mathbf{u}_k) \in \mathbb{Z} $.
\end{enumerate}
\begin{proof}
    For Property (1), we know that $\boldsymbol{\rho}_k \in \mathcal{P}_k$ and define $	\epsilon_{\rho_k}=\boldsymbol{\rho}^*_{k+1}-\boldsymbol{\rho}^*_k$. Then we can get
    \begin{equation*}
        \mathcal{P}_k \cap \Delta_{k+1}= \left\{\boldsymbol{\rho}_k \in \mathbb{R}^3 \mid \tilde{\mathbf{H}}_{\theta_k} \boldsymbol{\rho}_k \leq \tilde{\boldsymbol{\delta}}_{\rho_k}\right\}.
    \end{equation*}
    After that, we can get
    \begin{equation*}
        \tilde{\mathbf{H}}_{\rho_k} \boldsymbol{\rho}^*_{k+1}=\tilde{\mathbf{H}}_{\rho_k} \boldsymbol{\rho}^*_k+\tilde{\mathbf{H}}_{\rho_k} 	\epsilon_{\rho_k} \leq \tilde{\boldsymbol{\delta}}_{\rho_k}+d_\rho \boldsymbol{\gamma}.
    \end{equation*}
     Since by assumption $\boldsymbol{\rho}_{k+1} \in \mathcal{P}$, we can get $\boldsymbol{\rho}^*_{k+1} \in \mathbf{d}_{\varsigma}\left(\mathcal{P}_k \cap \Delta_{k+1}\right)$.

     For Property(2), since we know that 
     \begin{equation*}
         \mathcal{P}_{\varsigma \mid k+1}=\mathbf{d}_{\varsigma+1}\left(\mathcal{P}_k\right) \cap \mathcal{P}= \mathcal{P}_{\varsigma+1|k}.
     \end{equation*}
Then we have
\begin{equation*}
\begin{aligned}
&\mathbb{X}_{\varsigma+2 \mid k}^* \supseteq \mathbf{A}_\text{cl}(\boldsymbol{\rho}) \mathbb{X}_{\varsigma+1 \mid k}^* \oplus \mathbf{B}(\boldsymbol{\rho}) \mathbf{v}_{\varsigma+1 \mid k}^* \oplus \mathbb{W}\\
    \Rightarrow&\tilde{\mathbb{X}}_{\varsigma+1 \mid k+1} \supseteq \mathbf{A}_\text{cl}(\boldsymbol{\rho}) \tilde{\mathbb{X}}_{\varsigma \mid k+1} \oplus \mathbf{B}(\boldsymbol{\rho}) \tilde{\mathbf{v}}_{\varsigma \mid k+1} \oplus \mathbb{W}.
\end{aligned}
\end{equation*}
where $\tilde{\mathbf{v}}_{\varsigma \mid k+1}=\mathbf{v}_{\varsigma+1 \mid k}^*$ and $\varsigma \in \{0,N{-}2\}_\mathbb{N}$. Based on Assumption \ref{asmp::3} that there exist $\mathbb{X}_{N \mid k+1}$ satisfy the last set when $\varsigma = N{-}1$, that proves that $ \mathcal{I}_{v_s} \neq \{\varnothing\} $ during all time step.

Property (3) is the direct result of Section \ref{SetDynamics}.
\end{proof}

\section{Simulation results}
\label{sec::sim_results}
\begin{figure*}[!t]
\centering
\captionsetup{font=footnotesize}
\includegraphics[width=1\textwidth]{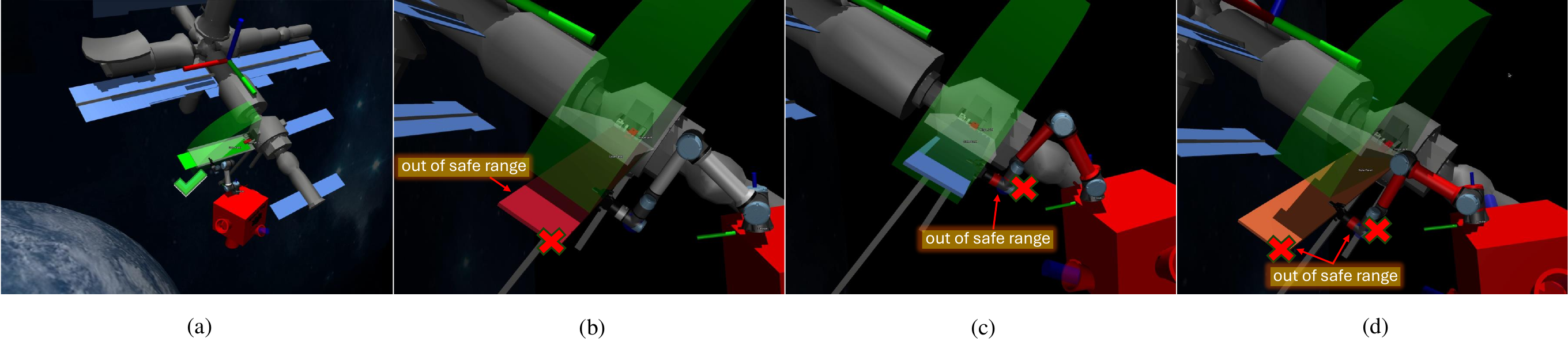}
\caption{Dislodging mission using a servicer which contains simulated on MuJoCo Platform in a freefall environment.}
\label{result::state_space_traj_compare}
\end{figure*}

\begin{figure*}[!t]
    \centering
    \captionsetup{font=footnotesize}
    \begin{subfigure}[c]{0.33\textwidth}
        \centering
        \vspace{-1.5mm}
        \captionsetup{font=footnotesize,,margin={0.7cm,0cm}}
        \includegraphics[width=1.09\linewidth]{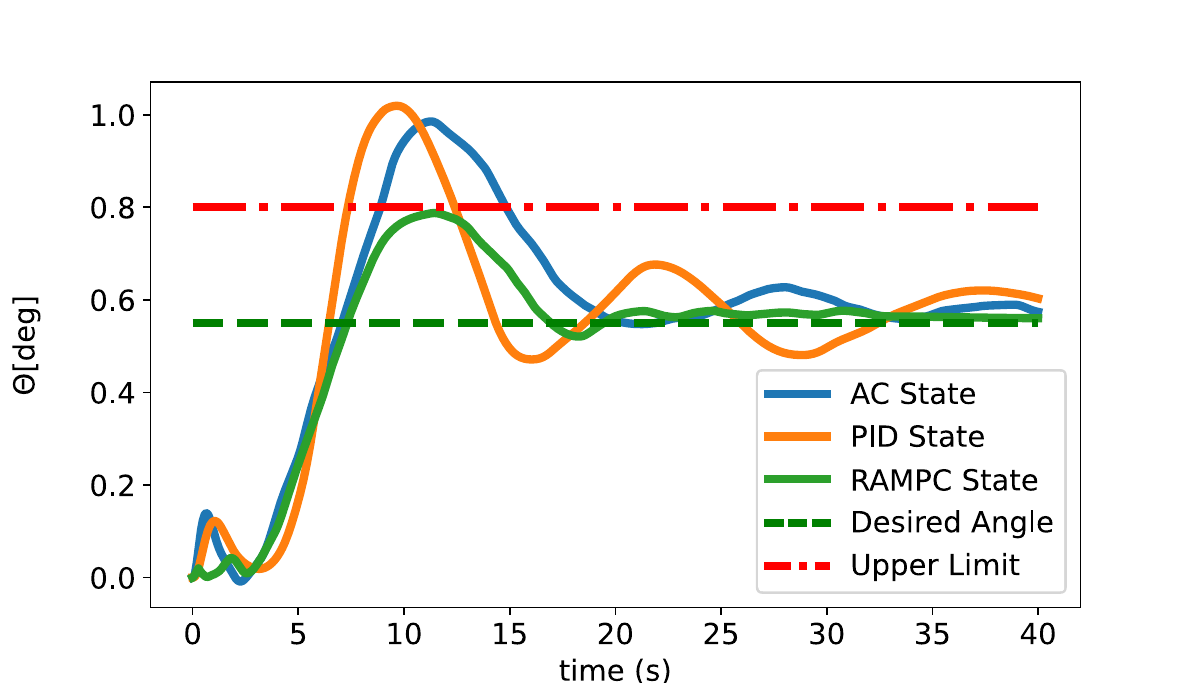}
        \subcaption{}
        \label{res::theta}
        \vspace{-3.2mm} 
        \includegraphics[width=1.09\linewidth]{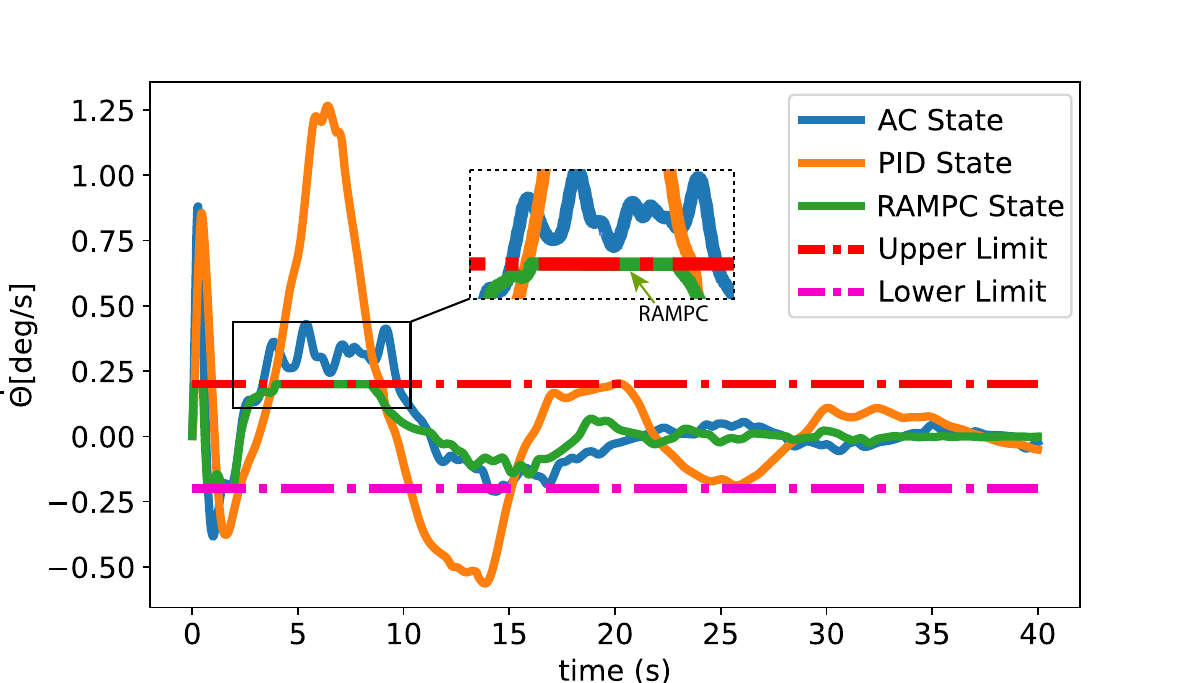}
        \subcaption{}
        \label{res::theta_dot_sim}
    \end{subfigure}
    \begin{subfigure}[c]{0.32\textwidth}
        \centering
        \vspace{0.4mm} 
        \captionsetup{font=footnotesize,,margin={0.7cm,0cm}}
        \includegraphics[width=1.1\linewidth]
        {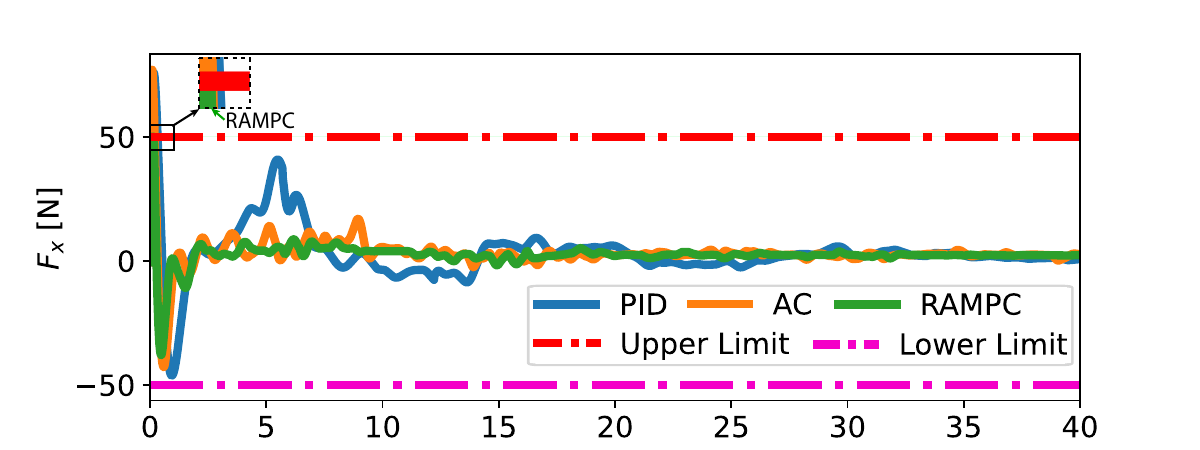}
        \includegraphics[width=1.1\linewidth]{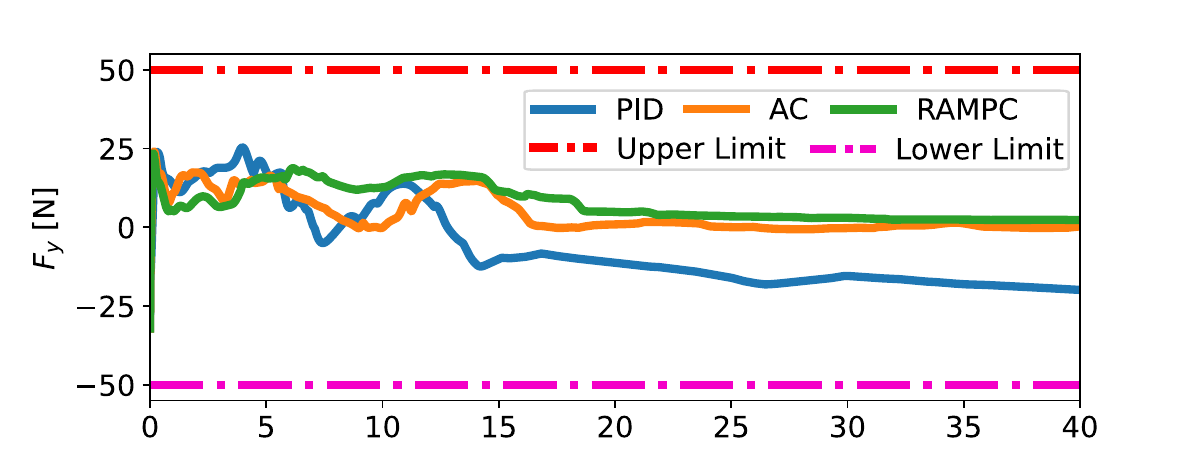}
        \includegraphics[width=1.1\linewidth]{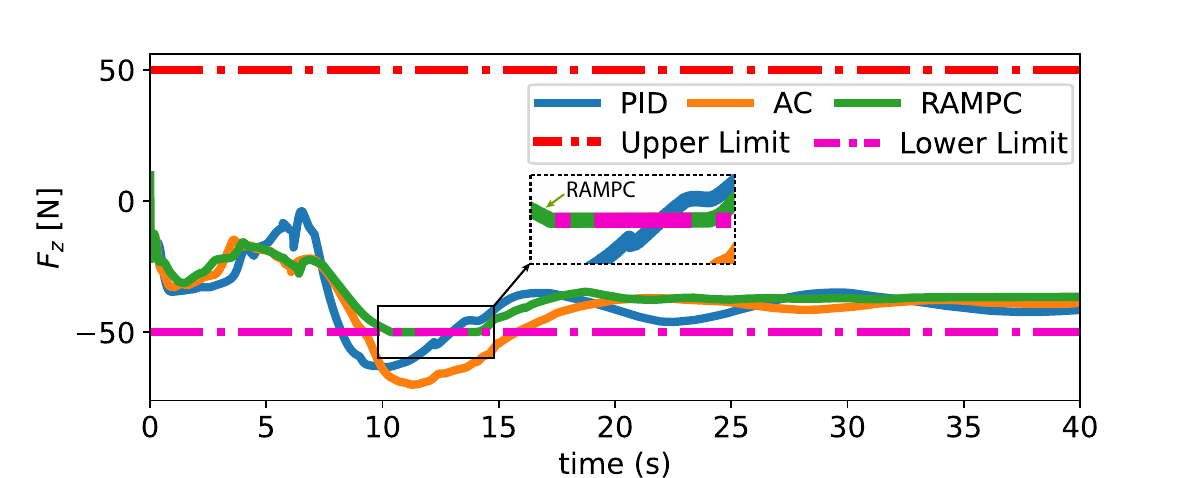}
        \subcaption{}
        \label{res::force}
    \end{subfigure}
    \begin{subfigure}[c]{0.32\textwidth}
        \centering
        \vspace{0mm} 
        \captionsetup{font=footnotesize,,margin={0.7cm,0cm}}
        \includegraphics[width=1.1\linewidth]{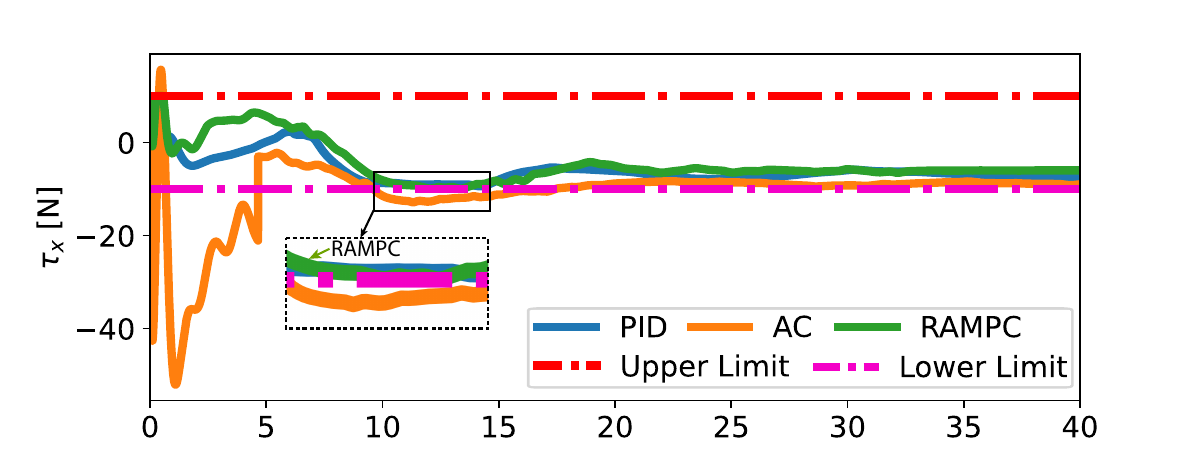}
        \includegraphics[width=1.1\linewidth]{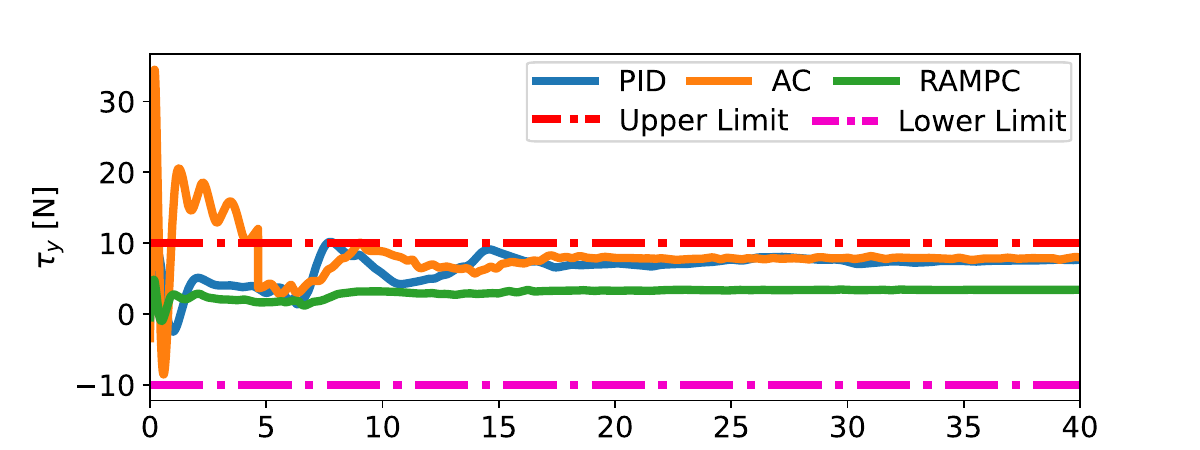}
        \includegraphics[width=1.1\linewidth]{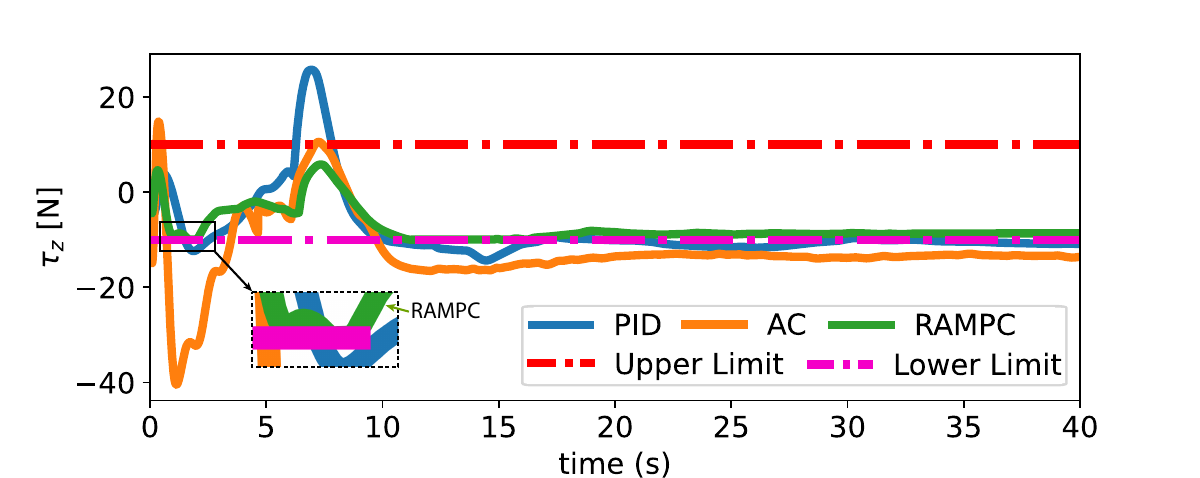}
        \subcaption{}
        \label{res::torque}
    \end{subfigure}
    \caption{Exploitation results for the evolution in both states of the client in which $\theta_k$ in (a) and  $\dot \theta_k$ in (b) and control input in which $\mathbf{f}_k$ in (c) and $\boldsymbol{\tau}_k$ in (d) compared to PID, adaptive control and robust adaptive MPC for 40 seconds.}
    \label{fig:four-images}
\end{figure*}

In this section, we formulate the discrete-time linear time-varying system of the client and the performance of our robust adaptive MPC is also presented in this section with the comparison with a baseline adaptive MPC algorithm in~\cite{aboudonia2024adaptive}. We choose the MuJoCo~\cite{todorov2012mujoco}, an advanced robotics physical platform which can fully emulate real space environment, as our simulation platform to emulate both client and servicer in a freefall environment. The parameters about the client are defined in Table.~\ref{tab:rampc_para}. Note that all parameters we selected in this section are to be physically reasonable and representative of our experimental hardware setup.

The uncertainty in the parameters is described by $ \mathcal{P} = \left\{\boldsymbol{\rho}_k \in \mathbb{R}^3\bigr|\: ||\boldsymbol{\rho}_k||_\infty \le 8 \right\}, $ with $ \boldsymbol{\rho}^*_k = [\mathcal{K}_k, \zeta_k, \mathcal{D}_k]^\top $. The disturbance set is $ \mathbb{W} = \left\{w\in \mathbb{R}^3 \bigr|\: ||w||_\infty \le 0.2 \right\} $ and note that our framework assumes a norm bounded disturbance rather than a specific frequency spectrum, providing robustness against any signal that remains within this set. The simulations employ box constraints, which are a simplified instance of our general polytopic formulation. This was a deliberate choice to isolate the performance benefits of the robust adaptation mechanism, as the baseline methods do not accommodate the more complex parameter-dependent constraints. The state and input constraints are described by 
\begin{align*}
	\mathbb{Z} &= \left\{(\mathbf{x},\mathbf{u}){\in} \mathbb{R}^{2 \times 6}\left|\: \begin{array}{rl}
	||\mathbf{x}[1]|| \le 0.7 &, ||\mathbf{x}[2]|| \le 0.2 \\
	\norm*{\mathbf{u}[1,..,3]} \le 50 &, \norm*{\mathbf{u}[4,..,6]} \le 10 \\ 
	\end{array}\right.  \right\}.
\end{align*}

The initial state of the Newton system is $\mathbf{x}_0 = [0.6,0.2]^\top$. We set up the sampling period parameter $h$ in Eq.~\ref{eq:parametric-plant} as $0.01s$ to maintain consistency for on the hardware feature and simulation platform. In both AMPC and our method, the state tube is constructed by translating and scaling the set $\mathbb{X}_0 = \left\{\mathbf{x}\in \mathbb{R}^2 \bigr|\: \norm*{\mathbf{x}[1]} \le 0.8, \norm*{\mathbf{x}[2]} \le 0.25 \right\}$. The bounded complexity update of $ \mathcal{P}_k$ is performed using $n_\theta = 45$ hyperplanes which are initially chosen as outer bounds of the set $\mathcal{P}$. The cost matrices are given as 
\begin{equation*}
\begin{array}{l l l}
\mathbf{Q} = \begin{bmatrix} 1.47 &  0 \\ 0 & 1.35 \end{bmatrix}, & \:  \mathbf{R} = \begin{bmatrix} 0.94 &  0 \\ 0 & 1.0 \end{bmatrix}, & \: \boldsymbol{\Upsilon} = \begin{bmatrix} 1.15 &  0 \\ 0 & 1.25 \end{bmatrix}.
\end{array} 
\end{equation*}
The pre-stabilizing gain used is 
\begin{equation*}
	\mathbf{K} = \begin{bmatrix}
	-0.73 & 0.45 \\ 0.29 & 0.1
	\end{bmatrix}.
\end{equation*}

The prediction horizon chosen is $N = 6$ time steps for both algorithms. Both of our scheme and AMPC are initialized at $ \hat{\boldsymbol{\rho}}_0 = [0.2,0.5, 2.0]^\top$. 

\begin{table*}[!htb]
    \centering
    \setlength{\tabcolsep}{6pt}  
    \captionsetup{font=footnotesize}
    \begin{tabular}{ccccccccc}
        \toprule
        Constraint Type & $\bar\theta$ & $\underline\theta$ & $\bar{\dot\theta}$ & $\underline{\dot\theta}$ & $\bar{\tau}_i$ & $\underline{\tau_i}$ & $\bar{f_i}$ & $\underline{f_i}$\\
        \midrule
        Experimental Value & 1.57\ rad & -1.57\,rad & 0.5\ rad/s & -0.5\ rad/s & 10\ N$\cdot$m & -10\ N$\cdot$m & 50\ N & -50\ N\\
        \textbf{Simulation Value} & \textbf{0.3\ rad} & \textbf{-1.2\,rad} & \textbf{0.2\ rad/s} & \textbf{-0.2\ rad/s} & \textbf{2\ N$\cdot$m} & \textbf{-2\ N$\cdot$m} & \textbf{20\ N} & \textbf{-20\ N}\\
        \bottomrule
    \end{tabular}
    \caption{Constraints for both \textit{Servicer} and \textit{Client} in experiments.}
    \label{tab:exp_constrain}
\end{table*}

\begin{figure*}[!b]
\centering
\captionsetup[sub]{font=footnotesize,oneside,margin={0.75cm,0cm}}
\begin{subfigure}{2.3in}
    \includegraphics[width=\textwidth]{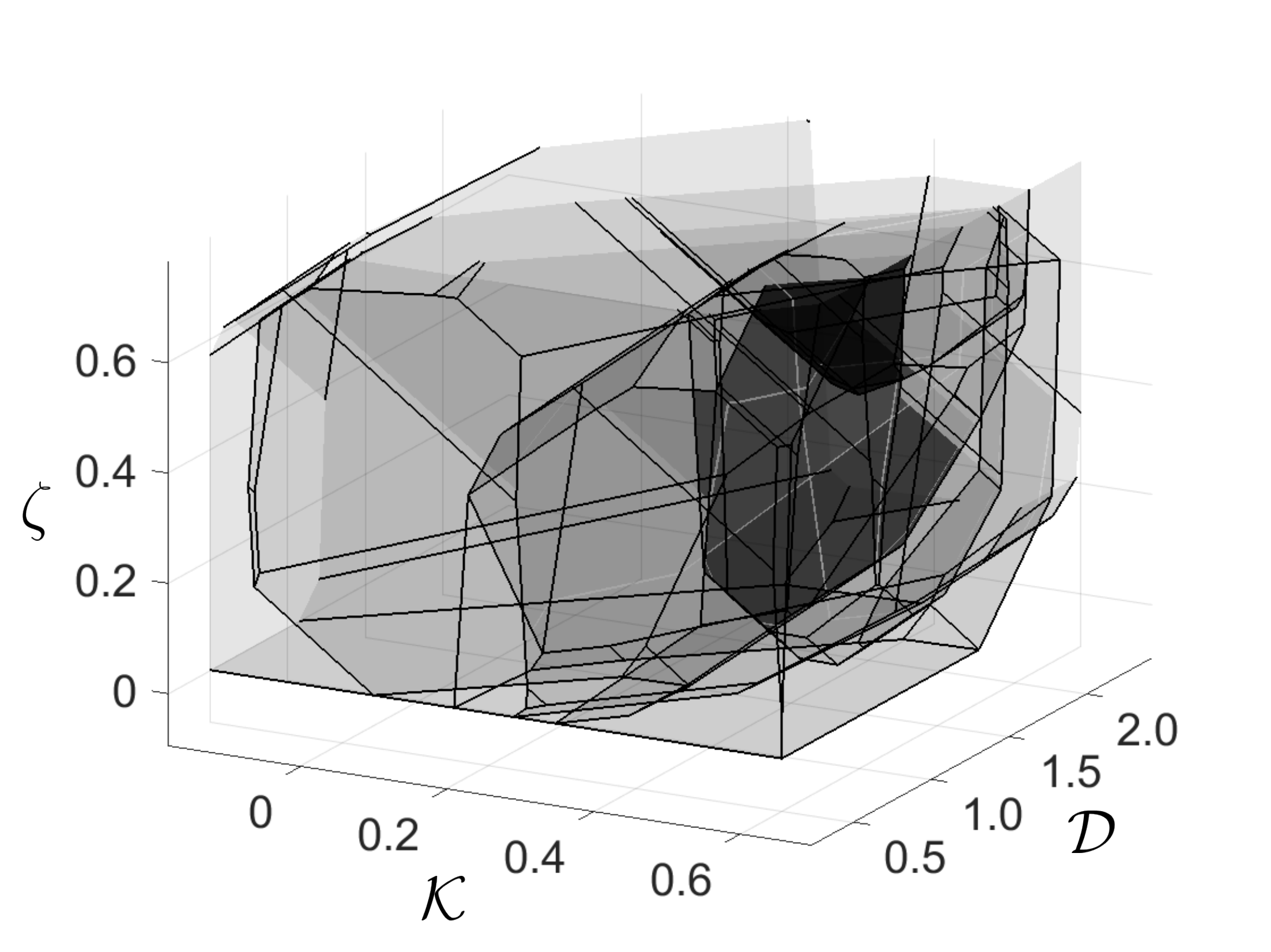}
    \caption{}
    \label{polygen::ac}
\end{subfigure}
\begin{subfigure}{2.3in}
    \includegraphics[width=\textwidth]{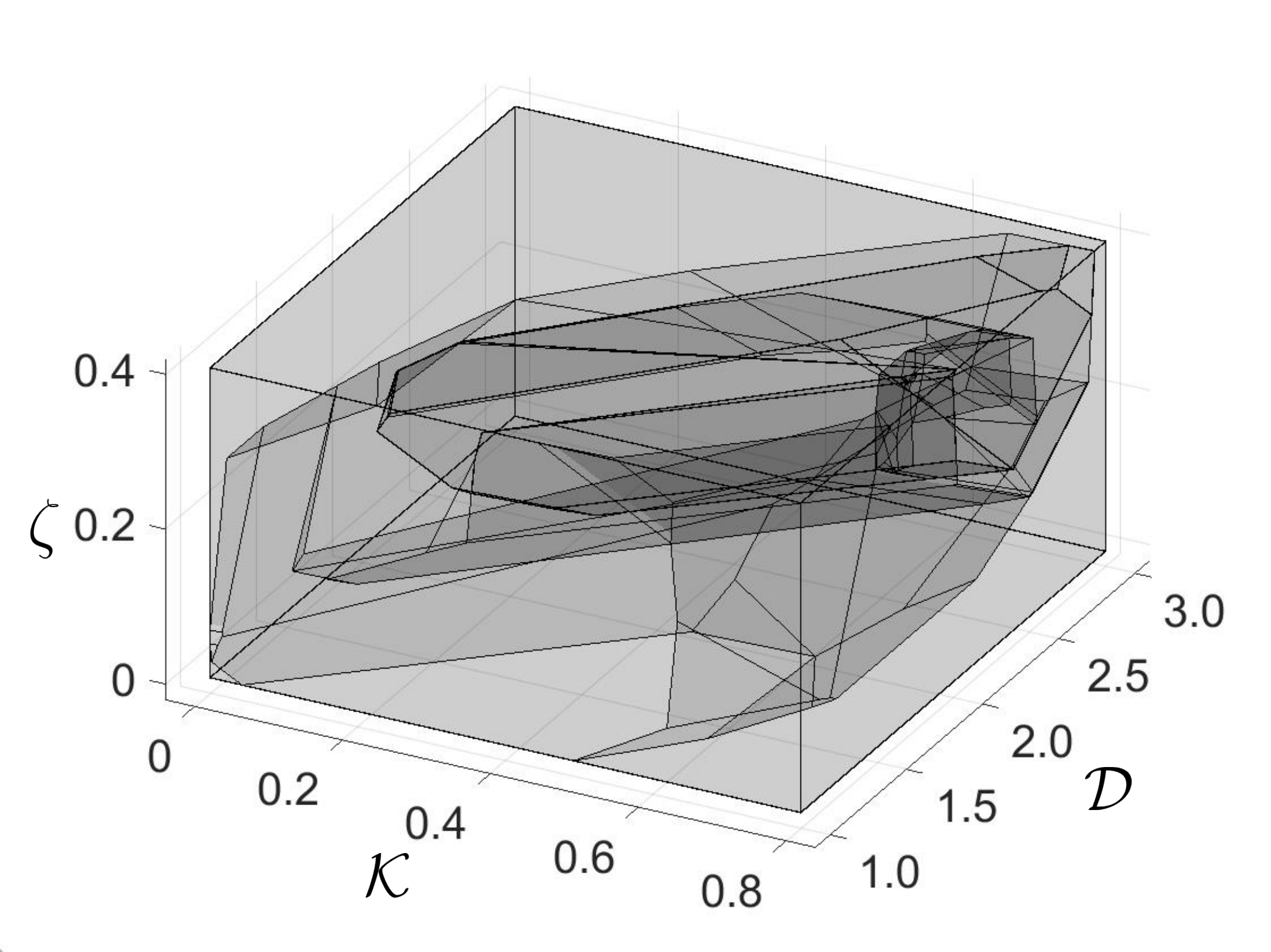}
    \caption{}
    \label{polygen::pampc}
\end{subfigure}
\begin{subfigure}{2.3in}
    \includegraphics[width=\textwidth]{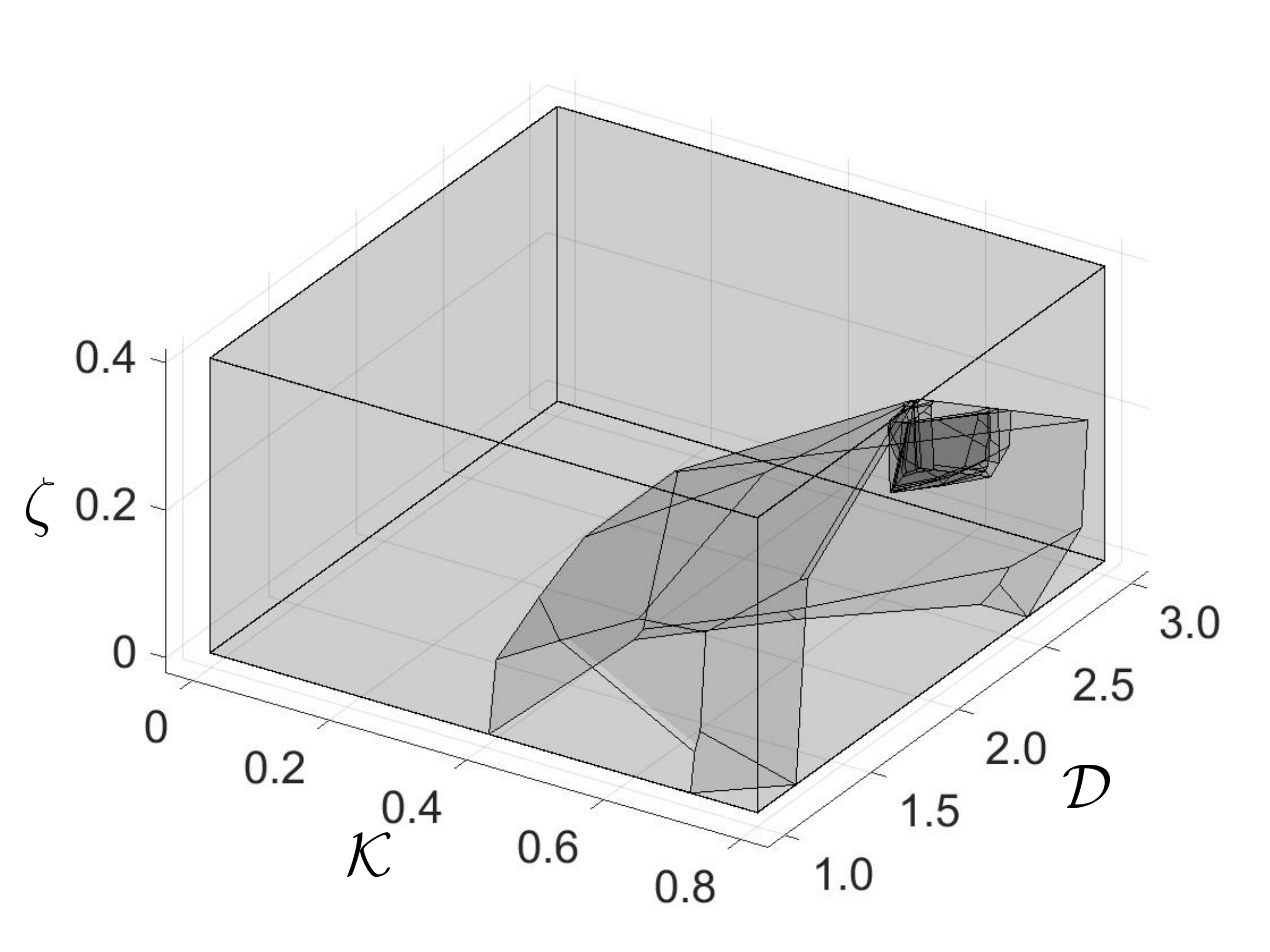}
    \caption{}
    \label{polygen::ours}
\end{subfigure}
\begin{subfigure}{2.3in}
    \includegraphics[width=\textwidth]{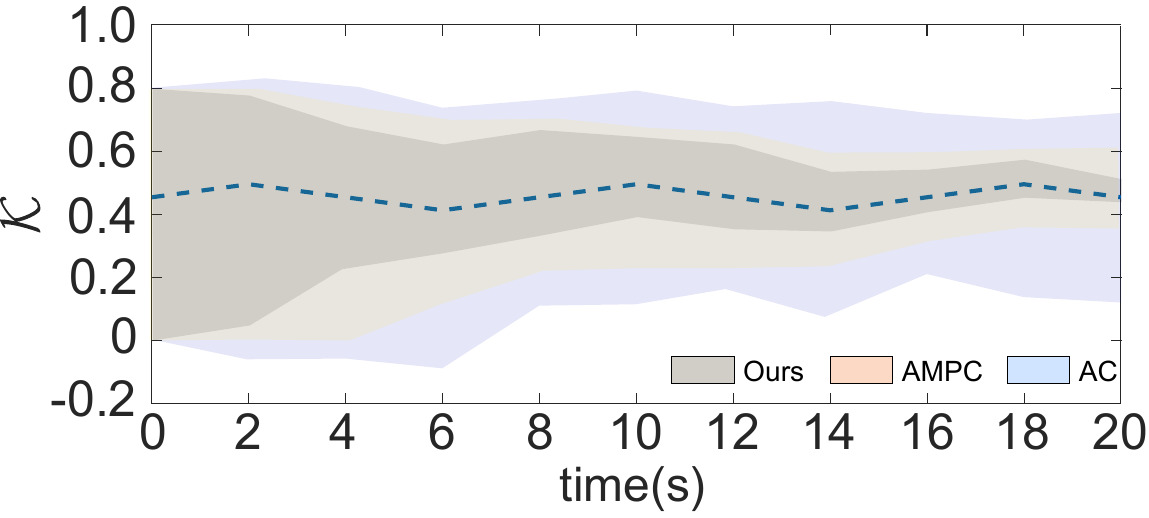}
    \caption{}
    \label{subfig::K}
\end{subfigure}
\begin{subfigure}{2.3in}
    \includegraphics[width=\textwidth]{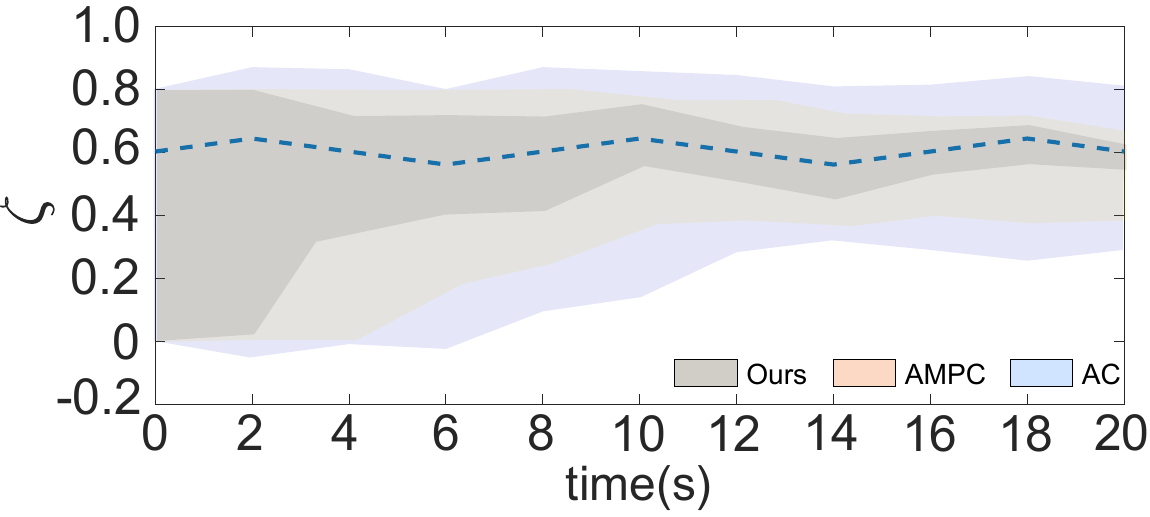}
    \caption{}
    \label{subfig::zeta}
\end{subfigure}
\begin{subfigure}{2.3in}
    \includegraphics[width=\textwidth]{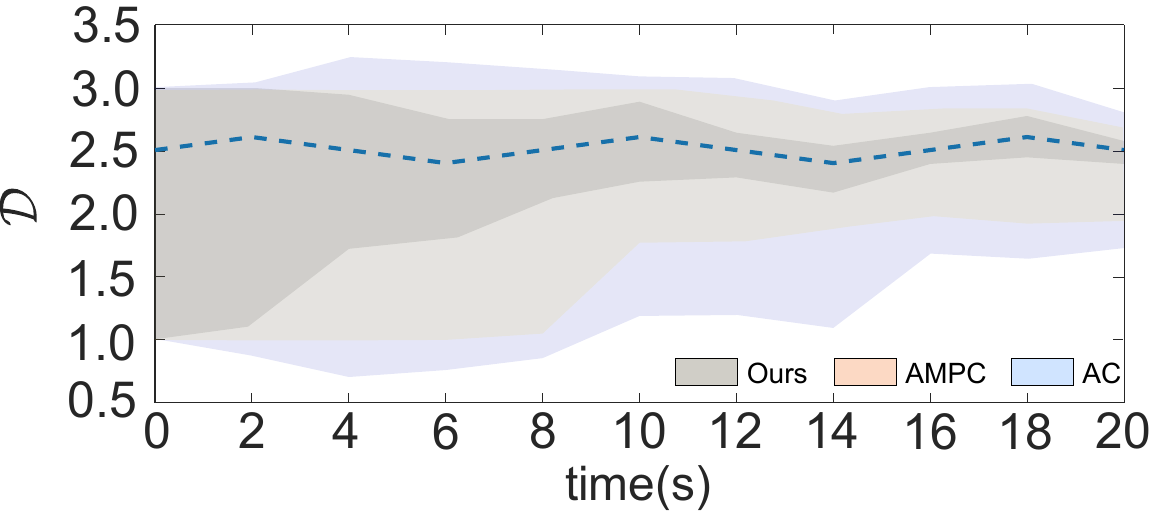}
    \caption{}
    \label{subfig::Dg_k}
\end{subfigure}
\caption{Exploration results for the evolution of the time-varying parameter membership set of the client during 20 seconds. (a) The polyhedron of the parameter update uses Adaptive control; (b) The polyhedron of the parameter update uses AMPC. (c) The polyhedron of the parameter update uses our Robust Adaptive MPC. (d). Comparison of time-varying parameter $\mathcal{K}$ estimation. (e). Comparison of time-varying parameter $\mathcal{\zeta}$ estimation. (f). Comparison of time-varying parameter $\mathcal{D}$ estimation. }
\label{result::polygen_for_compare}
\end{figure*}
Fig.~\ref{res::theta} and Fig.~\ref{res::theta_dot_sim} shows the results of the state variable based on our method compared with two baseline methods. Based on the comparison results, we can see clearly that our robust adaptive MPC is not only within the safe range but also converges to be stable faster than the PID and adaptive control method.

Fig.~\ref{res::force} and Fig.~\ref{res::torque} shows the results about the force and torque applied by the end-effector of the robot arm on the servicer, respectively. Based on the comparison results, we can see that the force applied based on our robust adaptive MPC algorithm is in the safe range and the magnitude of the growing speed at each time is flatter than the PID and adaptive methods which avoids the life reduction of the robot arm by excessive changes of its applied force on the EE.



Fig.~\ref{result::polygen_for_compare} presents a detailed comparison of the exploration performance, illustrating the evolution of the time-varying parameter estimation for our robust adaptive MPC (RAMPC) against the baseline adaptive control (AC) and AMPC methods. Fig.~\ref{polygen::ac} to Fig.~\ref{polygen::ours} depict the 3D evolution of the parameter membership set (polytope) over a 20s horizon. While all three controllers successfully reduce the size of the uncertainty polytope from its initial bounds, a stark difference in convergence efficacy is evident. The baseline AC in Fig.~\ref{result::polygen_for_compare} and AMPC in Fig.~\ref{polygen::pampc} controllers, which rely primarily on passive excitation, demonstrate a slow and less significant reduction, resulting in a large final uncertainty set. In striking contrast, our RAMPC in Fig.~\ref{polygen::ours} achieves a significantly more rapid and aggressive tightening of the parameter set, converging to a much smaller final polytope that closely encloses the true parameter trajectory, visualized as the dark inner line. This superior performance is a direct consequence of the novel dual-mode cost in \eqref{eq:MPCcost}, which explicitly penalizes the size of the predicted parameter uncertainty set. This term actively guides the controller to balance exploration to excite the system and reduce uncertainty with exploitation, thereby enhancing control performance. which is the feature that the baseline methods lack.

This quantitative advantage is further clarified in Fig.~\ref{subfig::K} to Fig.~\ref{subfig::Dg_k}, which plot the bounded estimates for each individual time-varying parameter $\mathcal{K}$, $\zeta$, and $\mathcal{D}$ against time. The shaded regions represent the bounds of the estimated parameter set at each time step, while the dashed line indicates the ground-truth, time-varying parameter value. Across all three plots, the bounds generated by our RAMPC converge more rapidly and remain significantly tighter around the true value compared to both the AMPC and AC baselines. This visualization confirms that our method not only achieves a faster rate of learning but also maintains a much smaller estimation error throughout the dislodging process. This robust parameter identification is critical, as it allows the controller to operate with less conservative bounds, directly improving control performance and safety, as validated in the state-tracking results shown in Fig.~\ref{fig:four-images}.

\section{Experiment Results}
\begin{figure*}[!t]
\centering
\captionsetup{font=footnotesize}
\includegraphics[width=0.95\textwidth]{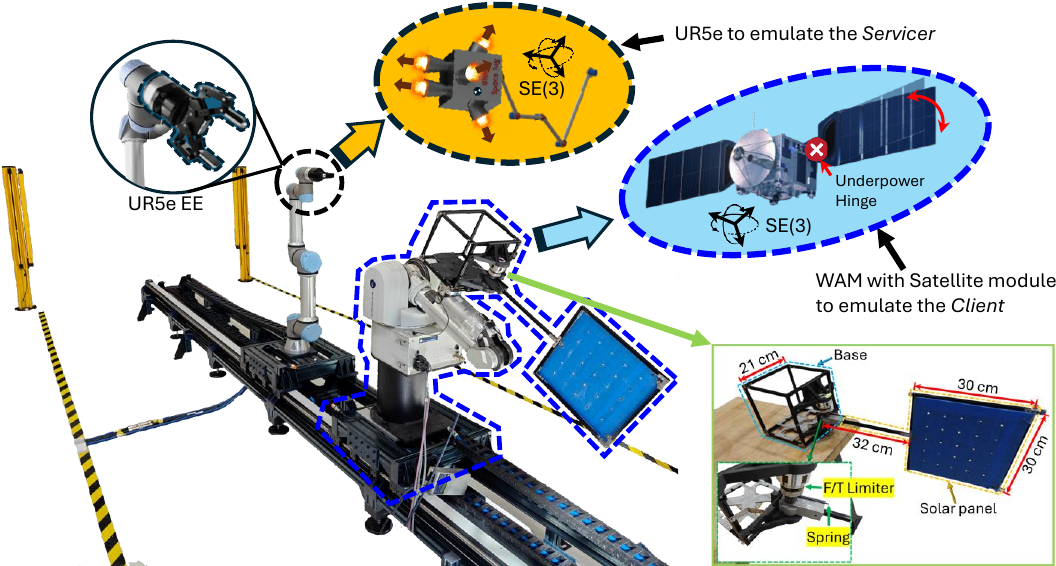}
\caption{The whole hardware setup for experiment. We use UR5e and WAM robot with gantry system to emulate the \textit{servicer} and \textit{client}, respectively. The detail of the satellite module design shown on the left-bottom side.}
\label{fig::hardware_whole}
\end{figure*}

\begin{figure*}[!htp]
\centering
\captionsetup{font=footnotesize}
\includegraphics[width=0.95\textwidth]{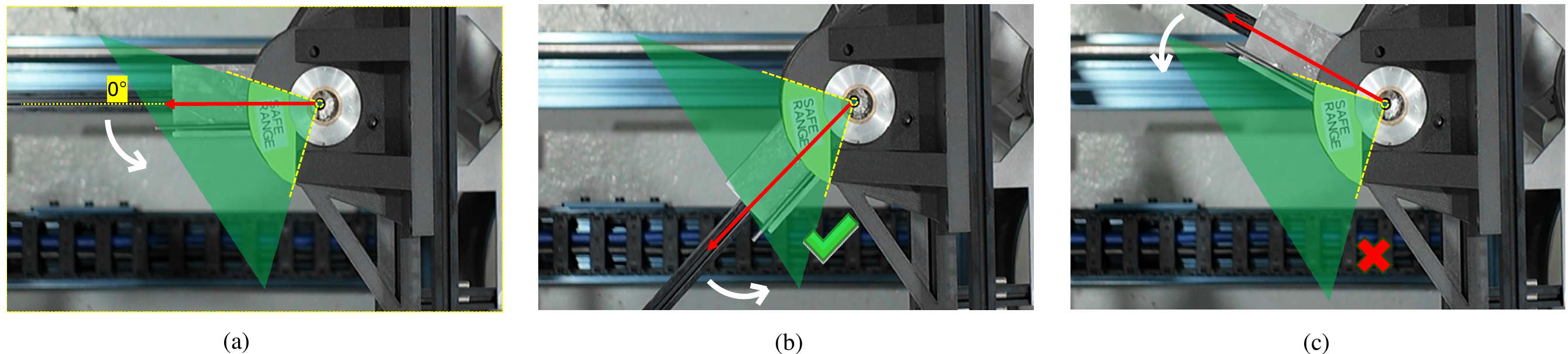}
\caption{Top-view of the hinge during the dislodging. The green region shows the safe dislodging range of the hinge's rotation. (a) shows the initial position of the hinge before the dislodging. (b) shows the hinge is in the safe range. (c) shows the hinge is out of the safe range.}
\label{fig::hinge_top_view}
\end{figure*}

\begin{figure}[!htbp]
    \centering
    \captionsetup{font=footnotesize}
    \begin{subfigure}[c]{0.48\textwidth}
        \centering
        \captionsetup{font=footnotesize,,margin={0.2cm,0cm}}
        \includegraphics[width=0.95\linewidth]{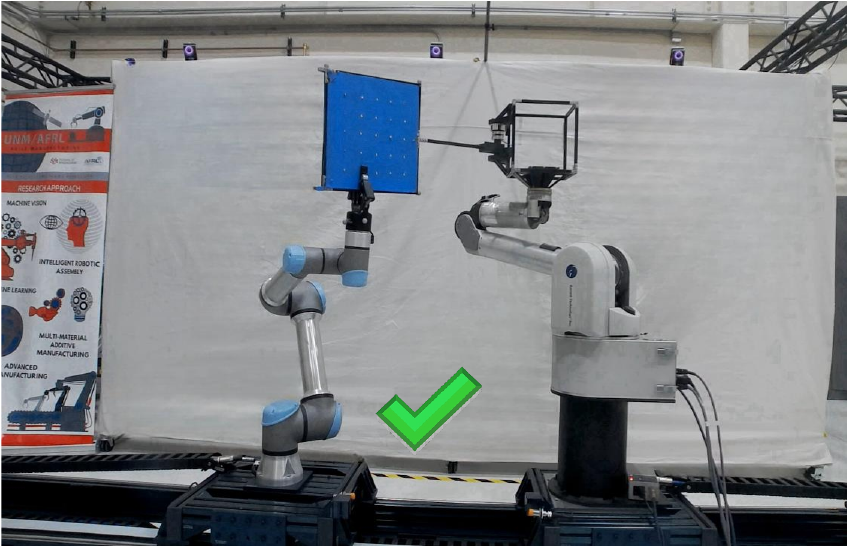}
        \label{fig::work_well}
        \subcaption{}
    \end{subfigure}
    \begin{subfigure}[c]{0.48\textwidth}
        \centering
        \captionsetup{font=footnotesize,,margin={0.2cm,0cm}}
        \includegraphics[width=0.95\linewidth]{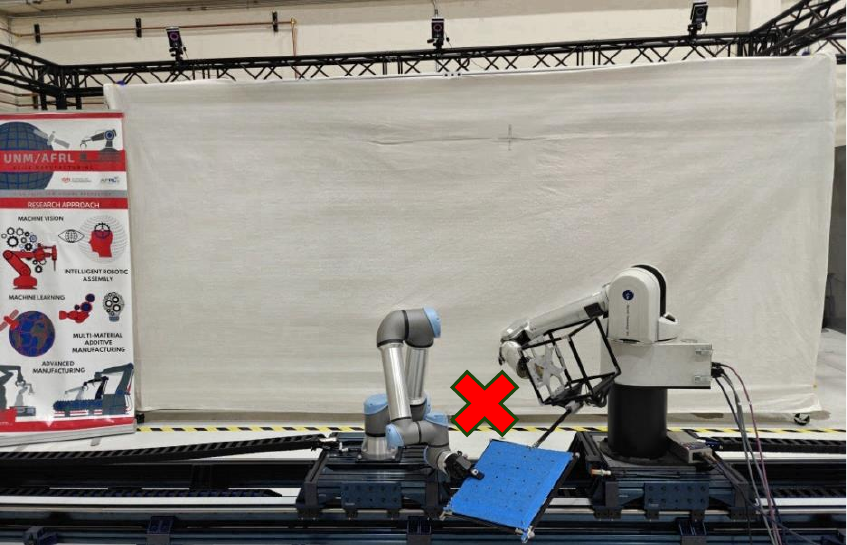}
        \label{fig::work_bad}
        \subcaption{}
    \end{subfigure}
    \caption{Hardware status during the dislodging. (a) shows the normal status during the dislodging that both \textit{Client} and \textit{Servicer} are in safe range. (b) shows the failure of dislodging that either is out of its safe range based on the hardware constraints in Table.II.}
    \label{fig::exp_well_bad}
\end{figure}

To prove the feasibility of our robust adaptive MPC in hardware, we also implement our robust adaptive MPC in our testbed compared with PID and adaptive control, as shown in Fig.~\ref{fig::hardware_whole}. We leverage the testbed in our lab to simulate the space dislodging environment considering the $1 g$ gravity based on the terrestrial laboratory. We use UR5e\protect\footnotemark[3] to emulate a \textit{Servicer} that can apply 6 DoF wrench in SE(3). Also, we use the WAM robot arm\protect\footnotemark[4] to hold the satellite module, the details of which are shown on the right-bottom side, to emulate a \textit{Client}, which is a free-flying spacecraft that can decouple the dynamics from the \textit{Servicer} during dislodging.  \footnotetext[3]{\href{https://www.universal-robots.com/products/ur5e/}{https://www.universal-robots.com/products/ur5e/}, by Universal Robot.}
\footnotetext[4]{\href{https://barrett.com/wam}{https://barrett.com/wam}, by Barrett Technology.} For both \textit{Client} and \textit{Servicer}, we set up two types of safety constraints in our algorithm as shown in Table.~\ref{tab:exp_constrain} based on the hardware limitation to ensure the dislodging can proceed well as shown in Fig.~\ref{fig::exp_well_bad}a; once either is out of its hardware constraints based on the value in Table.~\ref{tab:exp_constrain}, the system will stop to denote the potential risk occurrence during the dislodging as shown in Fig.~\ref{fig::exp_well_bad}b. Note that to compare and discover the performance with different methods, we set up a scheme in our program that if any variable's value shown in Table~\ref{tab:exp_constrain} is between the hardware constraints and software constraints, the robot system can still work but denote the potential damage will happen to simulate the real scenarios in space. The hinge's top-side view during dislodging is shown in Fig.~\ref{fig::hinge_top_view}. All the safe constraints we set up in our program are under the real constraints for both \textit{Client} and \textit{Servicer} to avoid any danger happening on the hardware.

Table~\ref{tab:metrics_compact} presents a quantitative evaluation of the proposed RAMPC against baseline controllers, providing compelling evidence of its superior performance across both simulation and hardware implementations. The metrics clearly demonstrate that the RAMPC framework achieves the lowest root mean square errors for both angle and velocity tracking, signifying enhanced control accuracy and robustness. Furthermore, our method exhibits a substantially lower constraint violation rate, a critical outcome that validates its capacity for ensuring operational safety during the complex dislodging maneuver. The controller also attains the fastest convergence times in both experimental settings, highlighting its superior efficiency. These empirical results collectively underscore the primary contribution of this work, affirming that the proposed robust adaptive MPC strategy effectively delivers a more accurate, safe, and efficient solution for the on orbit servicing task under significant parametric uncertainty.

\begin{figure}[!htp]
    \centering
    \includegraphics[width=0.95\linewidth]{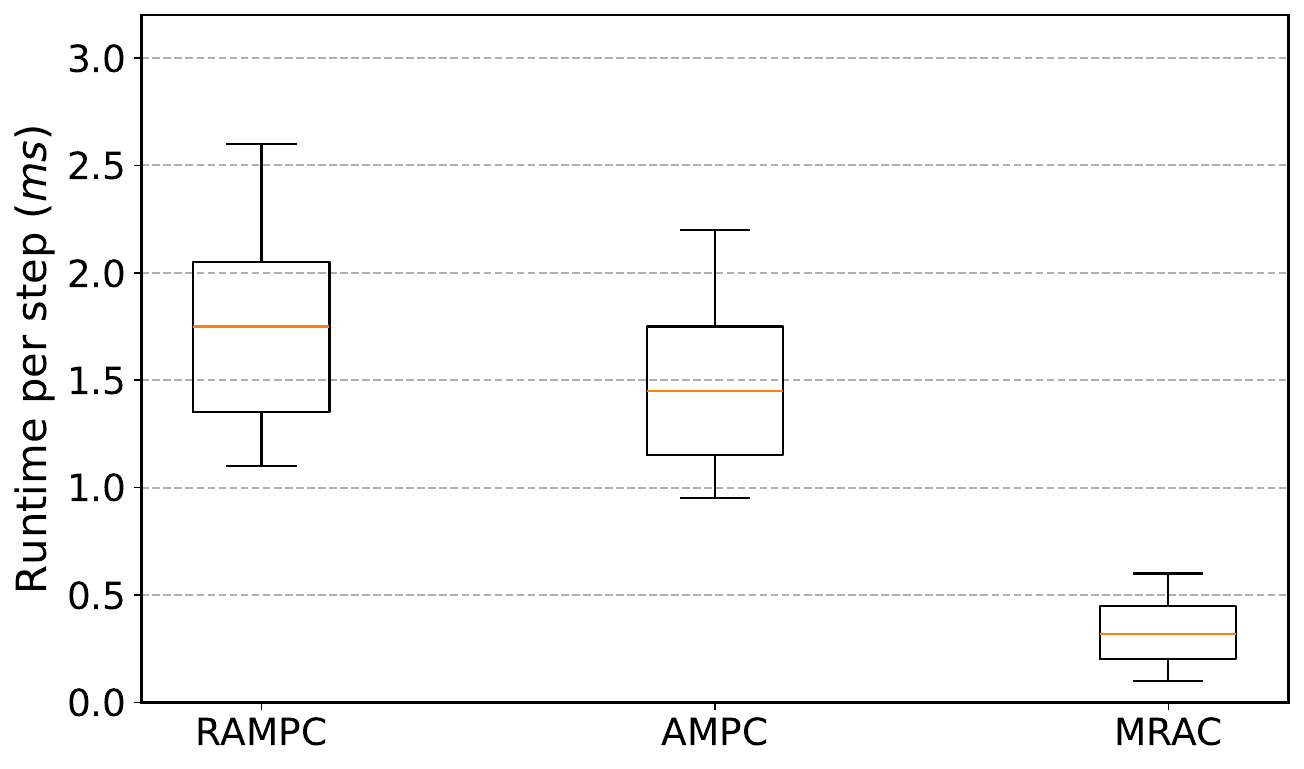}
    \caption{Runtime distributions in each step for RAMPC, AMPC, and MRAC at horizon six with warm starts. Boxes indicate the interquartile range and the center line denotes the median. Whiskers mark the lower and upper quantiles used for plotting and outliers are omitted. The vertical axis reports milliseconds per step. The control loop budget equals 10 ms.}
    \label{fig::runtime_1}
\end{figure}

Fig.~\ref{fig::runtime_1} presents a comparative analysis of the computational performance for the proposed Robust Adaptive Model Predictive Control RAMPC, a baseline Adaptive Model Predictive Control AMPC, and a Model Reference Adaptive Control MRAC scheme. The figure illustrates the runtime distribution in milliseconds for each control step, evaluated with a prediction horizon of six and employing warm starts to initialize the optimization. Both the RAMPC and AMPC methods exhibit comparable median computation times, which are observed to be well under two milliseconds. The interquartile ranges indicate consistent performance for both predictive controllers. In contrast, the non-predictive MRAC demonstrates substantially lower computational overhead, yet all three controllers comfortably satisfy the ten-millisecond control loop budget, confirming their suitability for real-time implementation.

\begin{figure}[!htbp]
    \centering
    \captionsetup{font=footnotesize}
    \begin{subfigure}[c]{0.48\textwidth}
        \centering
        \captionsetup{font=footnotesize,margin={0.7cm,0cm}}
        \includegraphics[width=0.8\linewidth]{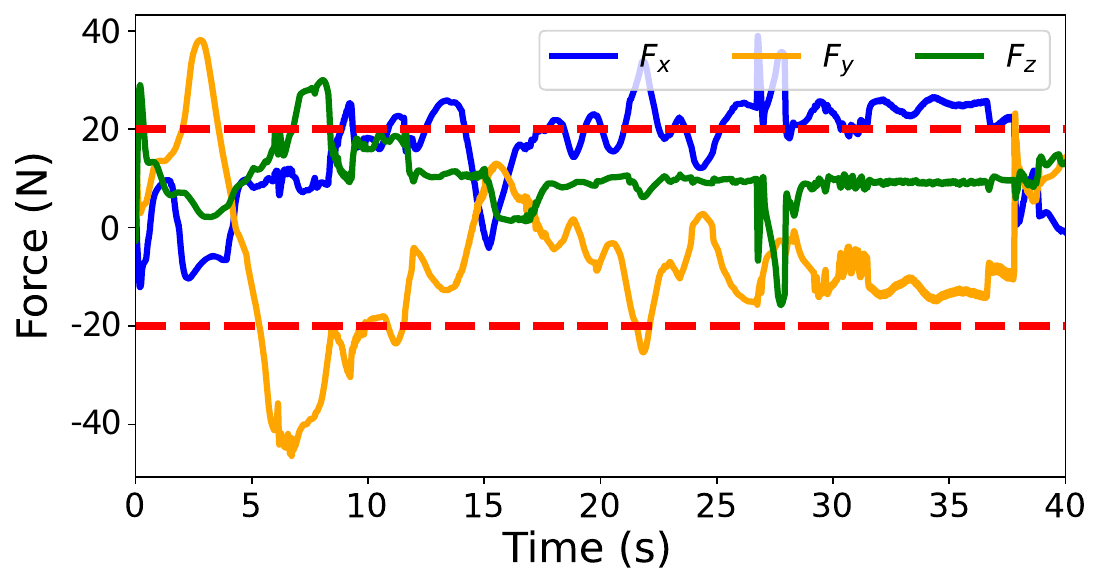}
        \label{res::force_pid}
        \subcaption{}
        \vspace{-1mm}
        \includegraphics[width=0.8\linewidth]{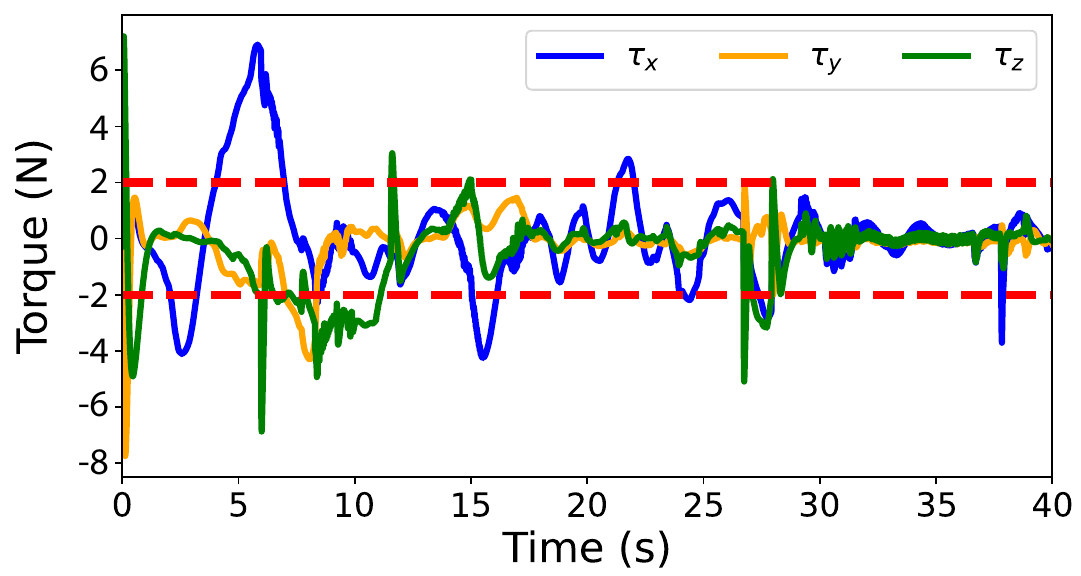}
        \label{res::tau_pid}
        \subcaption{}
    \end{subfigure}
    \caption{Force and torque applied by the EE of UR5e during the dislodging via PID.}
    \label{fig::pid_wrench}
\end{figure}

Fig.~\ref{fig::pid_wrench} shows the wrench applied by the \textit{Client} during the dislodging using PID control method. Based on the results, we can clearly see that the PID method can easily cause the \textit{Client} to exceed its safe range, which can damage the hardware.

\begin{figure}[!htbp]
\vspace{0.5cm}
    \centering
    \captionsetup{font=footnotesize}
    \begin{subfigure}[c]{0.48\textwidth}
        \centering
        \vspace{-1.5mm}
        \captionsetup{font=footnotesize,,margin={0.7cm,0cm}}
        \includegraphics[width=0.8\linewidth]{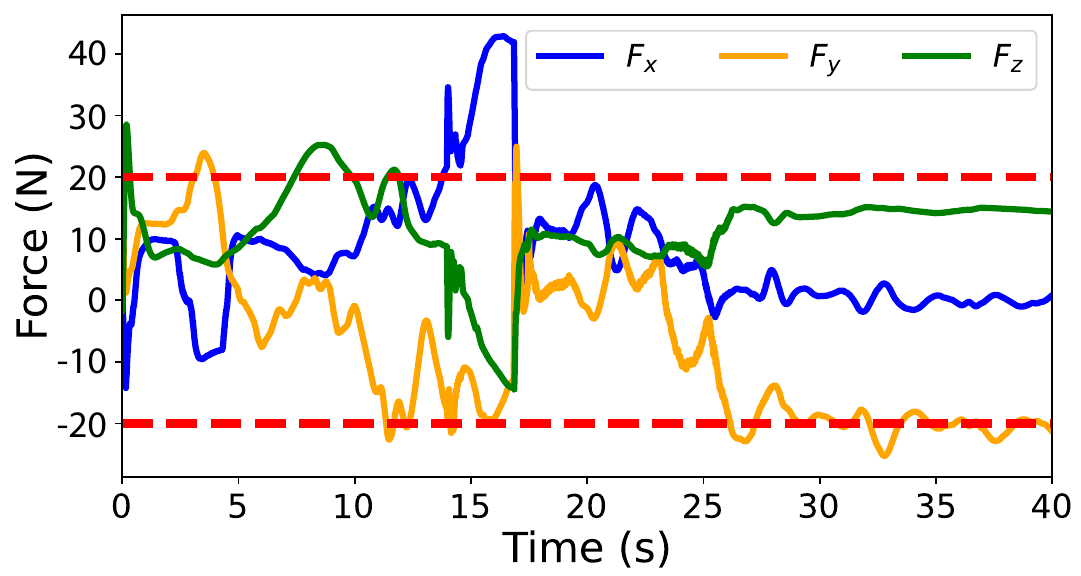}
        \label{res::force_ac}
        \subcaption{}
        \vspace{-1mm} 
        \includegraphics[width=0.8\linewidth]{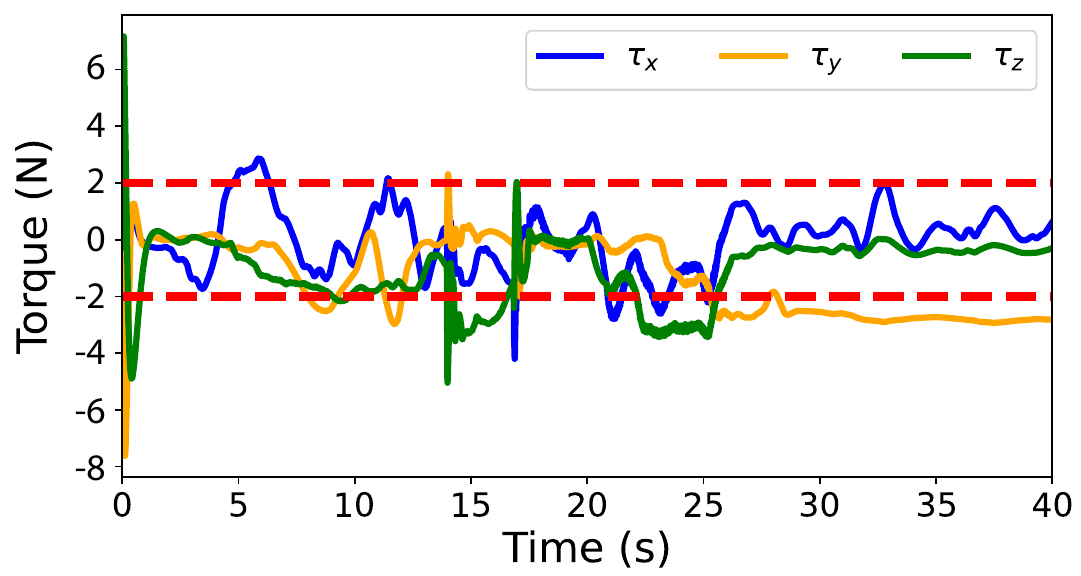}
        \subcaption{}
        \label{res::tau_ac}
    \end{subfigure}
    \caption{Force and torque applied by the EE of UR5e during the dislodging via Adaptive Control.}
    \label{fig::ac_wrench}
\end{figure}

Fig.~\ref{fig::ac_wrench} shows the wrench applied by the \textit{Client} during the dislodging via adaptive control. Based on the results, it's also straightforward to see that even though the adaptive control method shows a little bit better than PID, it still can make the \textit{Client} be out of its safe range, bring potential damage to the hardware.

\begin{table*}[!t]
\centering
\caption{Quantitative metrics from the plotted results under time varying uncertainty. Simulation uses zero gravity in MuJoCo with horizon six. Hardware uses the UR5e and WAM setup with software limits from Table II. Values are averages across trials estimated from the published figures.}
\setlength{\tabcolsep}{3.5pt}
\footnotesize
\begin{tabular}{lccc@{\hspace{10pt}}lccc}
\toprule
\multicolumn{4}{c}{Simulation} & \multicolumn{4}{c}{Hardware} \\
\cmidrule(lr){1-4}\cmidrule(lr){5-8}
Metric & RAMPC & AMPC & Adaptive & Metric & RAMPC & Adaptive & PID \\
\midrule
Angle RMSE [rad]                    & 0.003 & 0.004 & 0.005 & Angle RMSE [rad]                    & 0.320 & 0.380 & 0.430 \\
Velocity RMSE [rad per second]     & 0.003 & 0.004 & 0.004 & Velocity RMSE [rad per second]     & 0.080 & 0.100 & 0.120 \\
Violation rate [per cent of steps] & 0.04   & 0.5   & 1.5   & Violation rate [per cent of steps] & 0.2   & 1.0   & 3.0   \\
Convergence time [s]               & 12.0  & 14.0  & 18.0  & Convergence time [s]               & 12.0  & 18.0  & 22.0  \\
\bottomrule
\end{tabular}
\label{tab:metrics_compact}
\end{table*}

\begin{figure}[!htbp]
    \centering
    \captionsetup{font=footnotesize}
    \begin{subfigure}[c]{0.48\textwidth}
        \centering
        \vspace{-1.5mm}
        \captionsetup{font=footnotesize,,margin={0.7cm,0cm}}
        \includegraphics[width=0.8\linewidth]{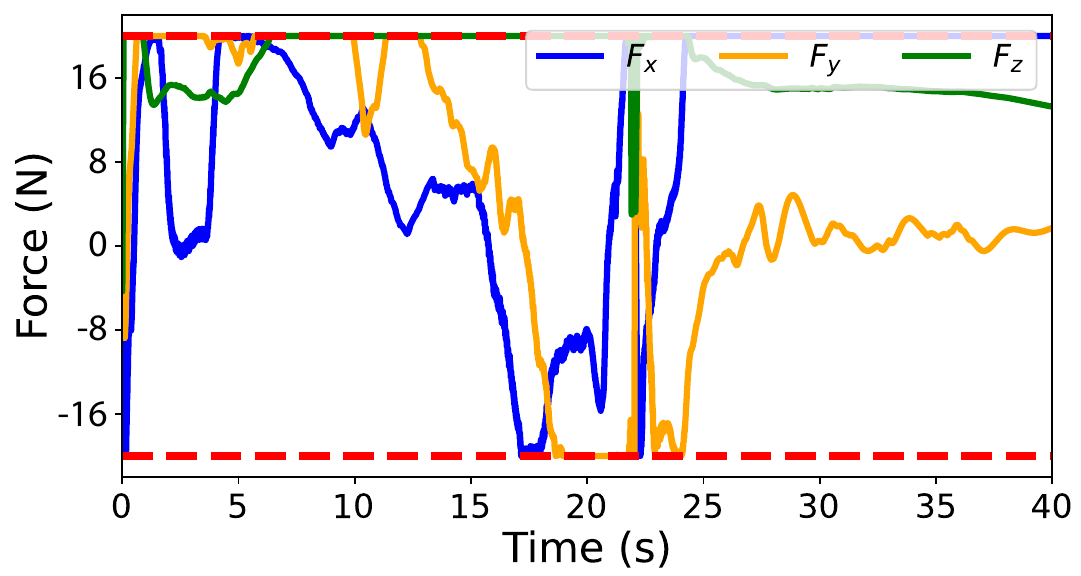}
        \label{res::force_mpc}
        \subcaption{}
        \vspace{-1mm} 
        \includegraphics[width=0.8\linewidth]{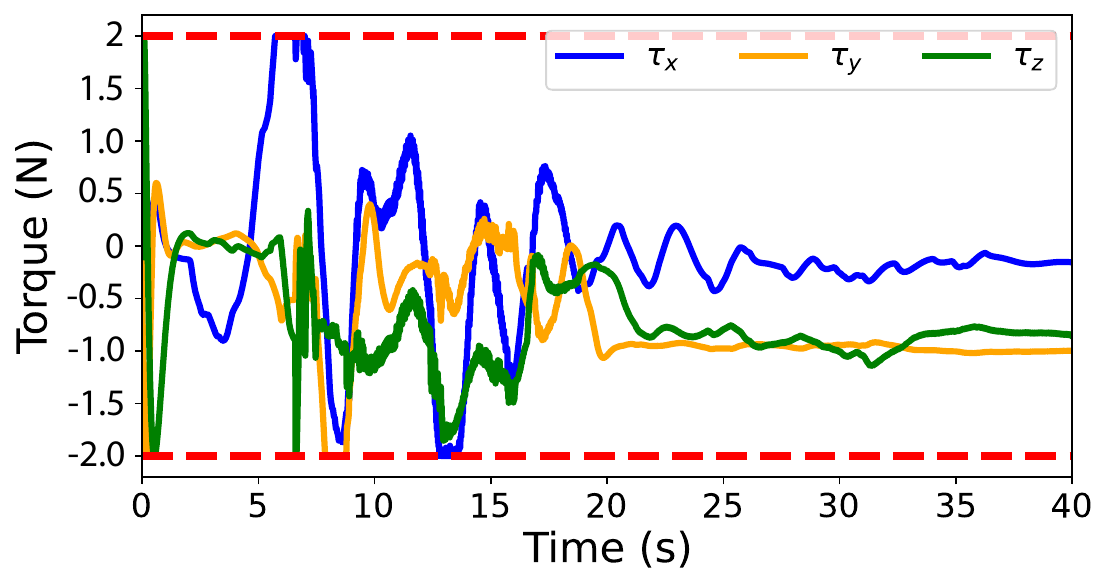}
        \subcaption{}
        \label{res::tau_mpc}
    \end{subfigure}
    \caption{Force and torque applied by the EE of UR5e during the dislodging via robust adaptive MPC.}
    \label{fig::mpc_wrench}
\end{figure}

Fig.~\ref{fig::mpc_wrench} shows the wrench applied by the \textit{Client} during the dislodging via our robust adaptive MPC. Based on the results, we can see that the adaptive control method ensures the wrench from the \textit{Client} is always within the constraint to maintain safety during dislodging.

\begin{figure}[!htbp]
    \centering
    \captionsetup{font=footnotesize}
    \begin{subfigure}[c]{0.48\textwidth}
        \centering
        \captionsetup{font=footnotesize,,margin={0.7cm,0cm}}\includegraphics[width=0.8\linewidth]{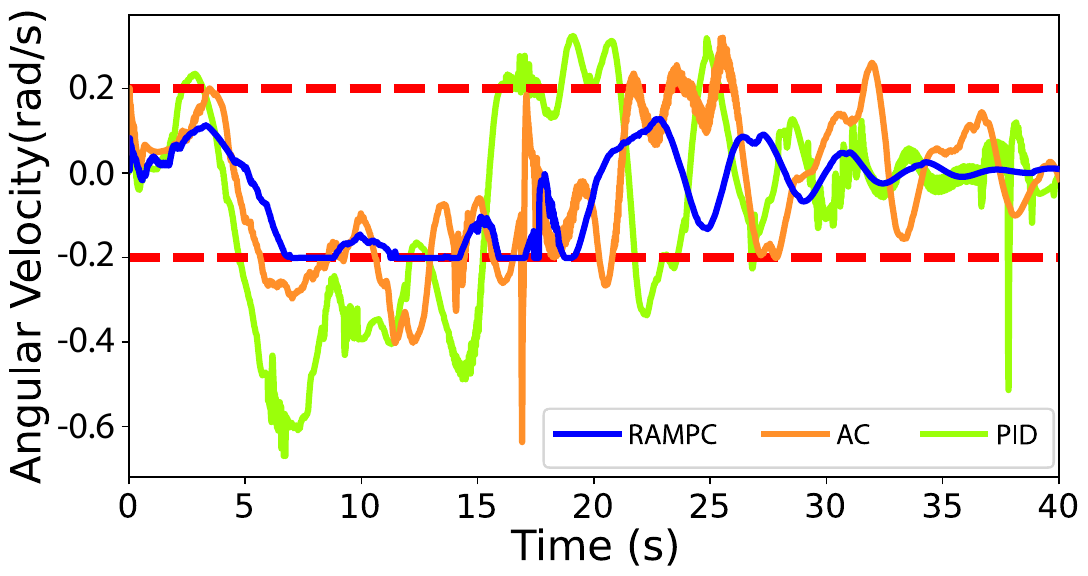}
        \subcaption{}
        \label{res::theta_dot} \includegraphics[width=0.8\linewidth]{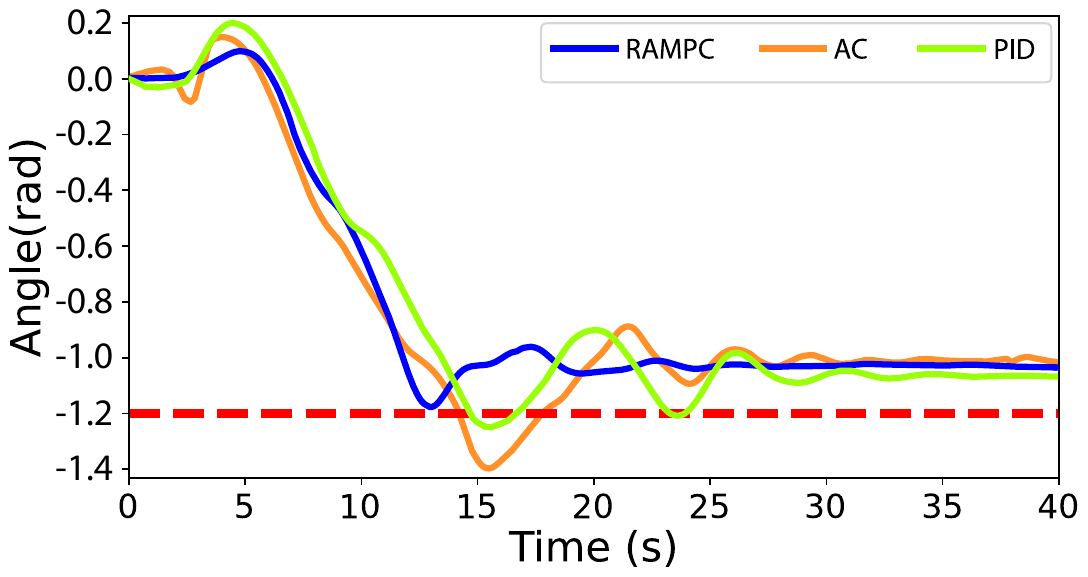}
        \subcaption{}
        \label{res::theta_dot} \includegraphics[width=0.8\linewidth]{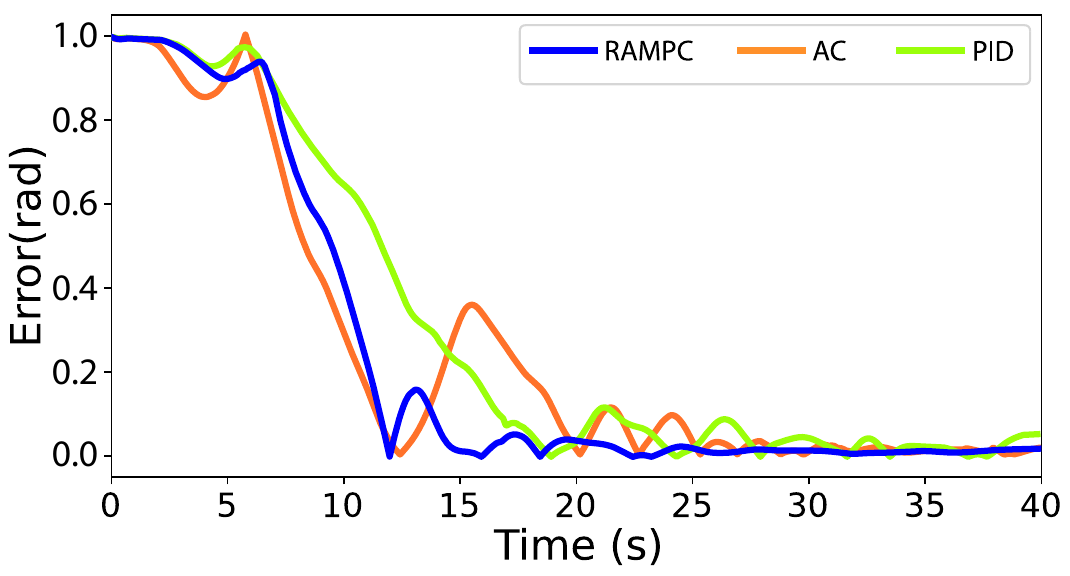}
        \subcaption{}
        \label{res::theta_dot}
    \end{subfigure}
    \caption{Angular states-angular velocity (a) and dislodged angle (b) and tracking error during the dislodging via three methods.}
    \label{fig::exp_error_vel}
\end{figure}

Fig.~\ref{fig::exp_error_vel} shows the comparison results for the angular velocity, angle position and tracking error during dislodging. We can clearly see that our robust adaptive MPC shows good performance in terms of both convergence speed and constraint limits.

\section{Conclusions}
In this paper, we presented a novel, robust adaptive MPC algorithm via set-membership in time-varying parameter estimation during the dislodging task. The algorithm employs an online set-membership identification technique to progressively minimize time-varying parameter uncertainty while utilizing a tube-based MPC strategy to guarantee robust compliance with system constraints. A predicted state tube is leveraged to account for the influence of future control inputs on the identification process while optimizing the anticipated worst-case cost. We compare our scheme with a state-of-the-art adaptive MPC algorithm and prove that our algorithm shows better performance in both parameter estimation and calculation speed during the manipulation process. In our future research, we will integrate related learning methods based on our robust adaptive MPC algorithm to improve the model accuracy with data-driven system identification.

\section*{Acknowledgements}\label{sec:acknowledge}
This material is based on research sponsored by Air Force Research Laboratory (AFRL) under agreements FA9453-18-2-0022 and FA9550-22-1-0093. Any opinions findings, and conclusions or recommendations expressed in this material are those of the authors and do not necessarily reflect the views of the United States Air Force. Also, we thank Bennett Russell for his help in setting up the experimental testbed.

\bibliographystyle{ieeetr}
\bibliography{main}

\end{document}